\documentclass[aps,rmp,reprint,amsmath,amssymb,graphicx,superscriptaddress,longbibliography]{revtex4-2}

\usepackage[dvips]{graphicx}
\usepackage{dcolumn}

\usepackage[dvips]{graphicx}

\usepackage{bm}

\usepackage {amsmath}
\DeclareMathOperator{\Tr}{Tr}
\usepackage{epstopdf}

\usepackage{graphicx,amssymb}
\usepackage{color}

\usepackage{float}
\usepackage{siunitx}
\usepackage{mathrsfs}

\begin{document}

\title{Quantum heat transport in condensed matter systems}

\author{Jukka P. Pekola}
\affiliation{Pico group, QTF Centre of Excellence, Department of Applied Physics, Aalto University School of Science, P.O. Box 13500, 00076 Aalto, Finland}
\affiliation{Moscow Institute of Physics and Technology, 141700 Dolgoprudny, Russia}
\author{Bayan Karimi} 
\affiliation{Pico group, QTF Centre of Excellence, Department of Applied Physics, Aalto University School of Science, P.O. Box 13500, 00076 Aalto, Finland}

\date{\today{}}

\begin{abstract}
In this Colloquium recent advances in the field of quantum heat transport
are reviewed. This topic has been investigated theoretically for several
decades, but only during the past twenty years have experiments on various mesoscopic systems become feasible. A summary of the theoretical basis for describing heat transport in one-dimensional channels is first provided. Then the main experimental investigations of quantized heat conductance due to phonons, photons, electrons, and anyons in such channels are presented. These experiments are important for understanding the fundamental processes that underly the concept of a heat conductance quantum for a single channel. Then an illustration on how one can control the quantum heat transport by means of electric and magnetic fields, and how such tunable heat currents can be useful in devices is given. This lays the basis for realizing various thermal device components such as quantum heat valves, rectifiers, heat engines, refrigerators, and calorimeters. Also of interest are fluctuations of quantum heat currents, both for fundamental reasons and for optimizing the most sensitive thermal detectors; at the end of the review the status of research on this intriguing topic is given.

\end{abstract}


\maketitle

\tableofcontents{}

\section{Introduction}\label{Introduction}
\label{intro}
In this review we present advances on fundamental aspects of thermal transport in the regime where quantum effects play an important role. Usually this means dealing with atomic scale structures or low temperatures, or combination of the two. The seminal theoretical work by~\textcite{Pendry1983} presented, almost 40 years ago, the important observation that a ballistic channel for any type of a carrier can transport heat at the rate given by the so-called quantum of thermal conductance $G_{\rm Q}$. During the present millenium the theoretical ideas have developed into a plethora of experiments in systems involving phonons, electrons, photons and recently also particles obeying fractional statistics. We give an overview of these experiments backed by the necessary theoretical framework. The question whether a channel is ballistic or not, and under what conditions, is interesting as such, but it has also more practical implications. If one can control the degree of ballisticity, i.e. the transmission coefficient of the channel, one can turn the heat current on and off. Such quantum heat switches, or heat valves as they are often called, will be discussed in this review as well. Furthermore heat current via a quantum element in an asymmetric structure can violate reciprocity in the sense that rectification of heat current becomes possible. The bulk of the review deals with the time average (mean) of the heat current. Yet, the fluctuations of this quantity are interesting and they provide a yardstick for the minimal detectable power and for the ultimate energy resolution of a thermal detector. We discuss such a noise and its implications in ultrasensitive detection. 

The review opens after the present introductory section by a theoretical discussion of thermoelectric transport in one-dimensional channels in Section~\ref{Thermoelectric transport}. In Section~\ref{Thermal conductance} we present the concept and method of how to measure heat currents in general. Section~\ref{Experimental setups} reviews the central elements of the experimental setups. After these general sections, we move to heat transport in different physical systems: phonons in Section~\ref{Phonons}, electrons and fractional charges in Section~\ref{Electrons}, and photons in Section~\ref{Photons} including some detailed theoretical discussion in the subsections. Section~\ref{Tunable quantum} presents experimental results on heat control by external fields. In Section~\ref{Photons Quantum} we move to the discussion of a superconducting qubit as a tunable element in quantum thermodynamics. Section~\ref{Heat current noise} gives an account of both theoretical expectations and experimental status on heat current noise and associated fluctuations of effective temperature. Section~\ref{Summary} concludes the review by a summary and outlook including prospects on useful thermal devices and some interesting physical questions related to quantum heat transport.
\begin{figure}
	\centering
	\includegraphics [width=\columnwidth] {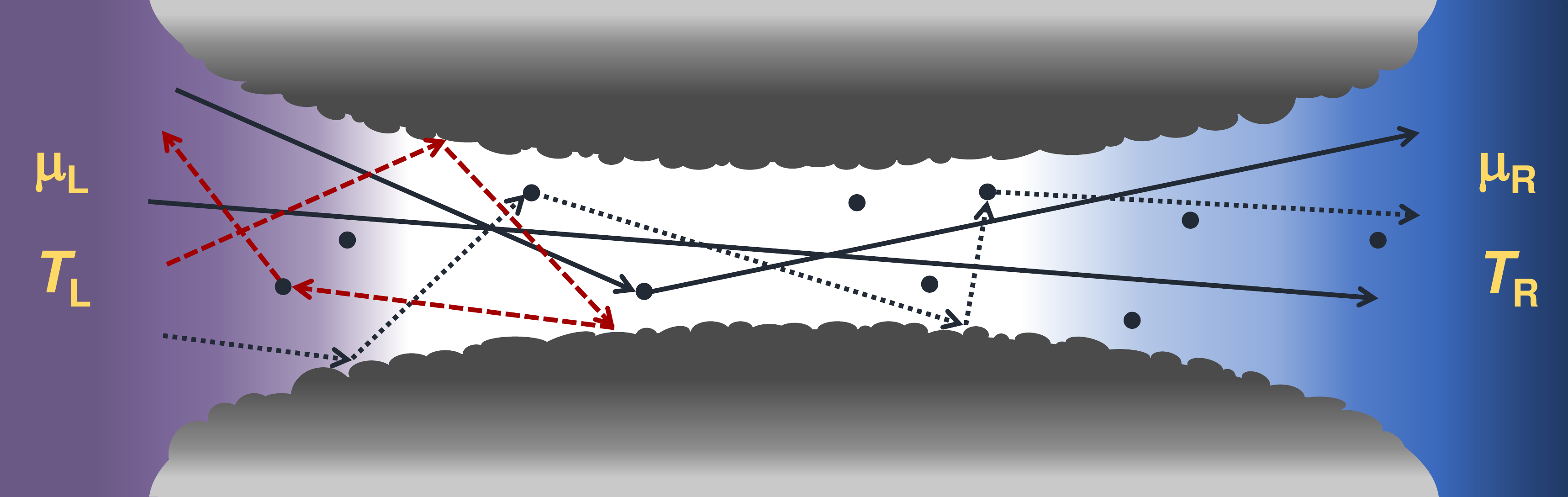}
	\caption{Artistic representation of a generic conductor between two reservoirs. Both particles and heat are transported through. Depending on the strength and type of scattering at the impurities (dots) and walls, one can have either ballistic or diffusive transport. Here $\mu_i$ and $T_i$, for $i=$L, R, are the chemical potential and temperature of each reservoir on the left and right, respectively.
		\label{one-channel}}
\end{figure}
\section{Thermoelectric transport in a one-dimensional (1D) channel}\label{Thermoelectric transport}
Consider two infinite reservoirs with temperature $T_i$ and chemical potential $\mu_i$ which are connected adiabatically via a conductor as shown schematically in Fig.~\ref{one-channel}. Here the subscripts $i$=L,~R represent the left and right, respectively. Based on Landauer theory~\cite{Landauer1981},~\cite{Sivan1986},~\cite{Butcher_1990} the charge and energy currents, $I$ and $\mathcal{J}$, between the two reservoirs (from L to R) are given for a 1D conductor by
\begin{eqnarray}\label{IQdot}
&&I=q\sum_{n}\int_{0}^{\infty}\frac{dk}{2\pi}v_n(k)(\vartheta_{\rm L}-\vartheta_{\rm R})\mathcal{T}_n(k)\nonumber\\&&\mathcal{J}=\sum_{n}\int_{0}^{\infty}\frac{dk}{2\pi}\varepsilon_n(k)v_n(k)(\vartheta_{\rm L}-\vartheta_{\rm R})\mathcal{T}_n(k),
\end{eqnarray} 
where $q$ is the particle charge, $\sum_{n}$ presents the sum over independent modes in the conductor, and $\varepsilon_n(k)$ and $v_n(k)$ indicate the energy and the velocity of the particles with wave vector $k$, respectively. $\mathcal{T}_n(k)$ indicates the particle transmission probability through the conductor via the channel; for ballistic transport $\mathcal{T}_n(k)\equiv 1$, and $\vartheta_{\rm L,R}$ represents the statistical distribution functions in each reservoir. Changing the variable from wave vector to energy via the definition of the velocity $v_n(k)=\frac{1}{\hbar}\frac{\partial \varepsilon_n(k)}{\partial k}$, we have 
\begin{eqnarray}\label{IQdot2}
&&I=\frac{q}{h}\sum_{n}\int_{\varepsilon(0)}^{\infty}~d\varepsilon[\vartheta_{\rm L}(\varepsilon)-\vartheta_{\rm R}(\varepsilon)]\mathcal{T}_n(\varepsilon)\nonumber\\&&\mathcal{J}=\frac{1}{h}\sum_{n}\int_{\varepsilon(0)}^{\infty}~d\varepsilon~\varepsilon[\vartheta_{\rm L}(\varepsilon)-\vartheta_{\rm R}(\varepsilon)]\mathcal{T}_n(\varepsilon),
\end{eqnarray} 
where $\varepsilon(0)\equiv\varepsilon$ for $k=0$. These equations constitute the basis of thermoelectrics, with linear response for electrical and thermal conductance and for Seebeck and Peltier coefficients. 

Now we solve analytically these equations for a ballistic contact $\mathcal{T}_n(\varepsilon)\equiv 1$ with the most common carriers, that is fermions and bosons. For fermions $\vartheta_i(\varepsilon)\equiv f_i(\varepsilon-\mu_i)=1/(1+e^{\beta_i(\varepsilon-\mu_i)})$ is the Fermi distribution function for each reservoir, with inverse temperature $\beta_i=1/(k_{\rm B}T_i)$. Note that we have taken the Fermi energy as the zero of $\varepsilon$, meaning that $\varepsilon(0)\rightarrow -\infty$. In this case at temperature $T$, with only chemical potential difference $eV$ across the contact, the charge current is 
\begin{eqnarray}\label{I1}
I=N\frac{e}{h}\int_{-\infty}^{\infty}d\varepsilon[f(\varepsilon)-f(\varepsilon-eV)]=N\frac{e^2V}{h}.
\end{eqnarray}
Here $N$ replacing the sum represents the number of current carrying modes in the conductor with $q\equiv e$. The electrical conductance $G=dI/dV$ is then
\begin{equation}\label{G1}
G=Ne^2/h,
\end{equation}
which is the famous quantization of electrical conductance. The thermal conductance for fermions can be obtained from the heat flux $\dot{Q}=\mathcal{J}$ when both reservoirs have the same chemical potential. The heat current across the ballistic contact is then
\begin{eqnarray}\label{Qdot1}
\dot{Q}=\frac{1}{h}\sum_{n}\int_{-\infty}^{\infty}d\varepsilon~\varepsilon[f_{\rm L}(\varepsilon)-f_{\rm R}(\varepsilon)].
\end{eqnarray}
The subtle differences between energy and heat currents are briefly discussed in Section~\ref{Experimental Heat}. In this review we focus mainly on thermal conductance in equilibrium, $T_{\rm L}=T_{\rm R}\equiv T$, i.e. on $G_{\rm th}(T)\equiv d\dot Q/dT_{\rm L}|_{T}$. The thermal conductance is then
\begin{eqnarray}\label{Gth1}
G_{\rm th}^{(f)}&&=N\frac{1}{h}\frac{1}{k_{\rm B}T^2}\int_{-\infty}^{\infty}d\varepsilon~\varepsilon^2~f(\varepsilon)[1-f(\varepsilon)]\nonumber\\&&=N\frac{\pi^2k_{\rm B}^2}{3h}T\equiv NG_{\rm Q},
\end{eqnarray}
where the superscript $(f)$ stands for fermions and 
\begin{equation}\label{GQ}
G_{\rm Q}\equiv \frac{\pi^2k_{\rm B}^2}{3h}T
\end{equation}
is the thermal conductance quantum. The ratio of the thermal and electrical conductances satisfies the Wiedemann-Franz law $G_{\rm th}^{(f)}/G=\mathcal{L}T$, where the Lorenz number is $\mathcal{L}=\frac{\pi^2k_{\rm B}^2}{3e^2}$~\cite{Ashcroft1976}.\\

We obtain the thermal conductance for bosons, $G_{\rm th}^{(b)}$, with the same procedure but with the distribution function $\vartheta_{\rm R,L}(\varepsilon)\equiv n_{\rm R,L}(\varepsilon)=1/(e^{\beta_{\rm R,L}\varepsilon}-1)$ in Eq.~\eqref{IQdot2}:
\begin{eqnarray}\label{Gthb}
G_{\rm th}^{(b)}=\frac{\hbar^2}{2\pi k_{\rm B}T^2}\sum_{n}\int_{0}^{\infty}d\omega\frac{\omega^2e^{\beta \hbar\omega}}{(e^{\beta \hbar\omega}-1)^2}\mathcal{T}_n(\omega).
\end{eqnarray}
Here $\varepsilon=\hbar\omega$ is the energy of each boson. For a single fully transmitting channel $\mathcal{T}_n(\omega)=1$, we then obtain again
\begin{equation}\label{Gthb2}
G_{\rm th}^{(b)}=G_{\rm Q}.
\end{equation}

Fermions and bosons form naturally the playground for most experimental realizations in the quantum regime. Yet the result above for a ballistic channel, $G_{\rm th}=G_{\rm Q}$, is far more general. As demonstrated in~\cite{Rego1999,Blencowe2000}, this expression is invariant even if one introduces carriers with arbitrary fractional exclusion statistics~\cite{Wu1994}. A few years back,~\cite{Banerjee2017} experimented on a fractional quantum Hall system addressing this interesting universality of thermal conductance quantum for anyons.  

\section{Thermal conductance - measurement aspects}\label{Thermal conductance}
\subsection{Principles of measuring heat currents}\label{Thermal Principles}
\begin{figure}
	\centering
	\includegraphics [width=\columnwidth] {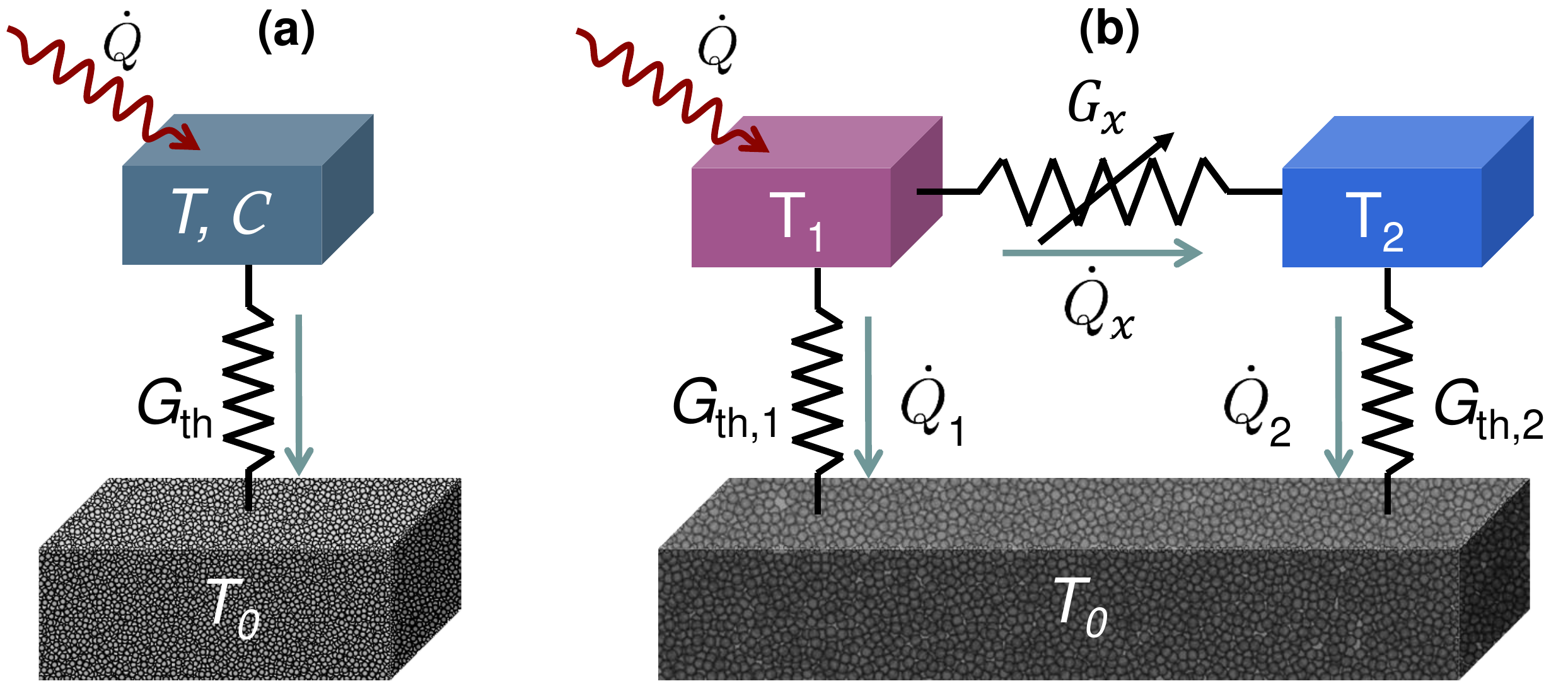}
	\caption{Thermal models. (a) A finite-sized reservoir at temperature $T$ and of heat capacity $\mathcal{C}$ coupled to a heat bath at fixed temperature $T_0$ via a heat link with thermal conductance $G_{\rm th}$. The absorbed heat current $\dot{Q}$ creates a temperature difference. (b) Two finite-sized absorbers each coupled to the heat bath and to each other via a potentially tunable thermal conductance $G_x$ with the associated heat current $\dot{Q}_x$ of the system under study.
		\label{epset}}
\end{figure}

For determining thermal conductance one needs in general a measurement of local temperature. Suppose an absorber, like the one in Fig.~\ref{epset}(a) is heated at a constant power $\dot Q$. By continuity, the relation between $\dot Q$ and temperature $T$ of the absorber with respect to the bath temperature $T_0$ can often be written as 
\begin{equation} \label{M1}
	\dot Q = \mathcal{K} (T^n-T_0^n),
\end{equation}
where $\mathcal{K}$ and $n$ are constants characteristic to the absorber and the process of thermalization. For instance, for the most common process in metals, coupling of absorber electrons to the phonon bath, the standard expression is $\dot Q = \Sigma \mathcal V (T^5-T_0^5)$~\cite{Roukes1985,Wellstood1994,Gantmakher1974,Schwab2000,Wang2019}, where $\Sigma$ is a material specific parameter and $\mathcal V$ the volume of the absorber. It is often the case that the temperature difference $\delta T \equiv T-T_0$ is small, $|\delta T/T| \ll 1$, and we can linearize Eq. \eqref{M1} into
\begin{equation} \label{M2}
	\dot Q = G_{\rm th} \delta T,
\end{equation} 
where $G_{\rm th} = n \mathcal{K}T_0^{n-1}$ is the thermal conductance between the absorber and the bath. For the electron-phonon coupling above, we then have $G_{\rm th}^{(\rm ep)} = 5\Sigma  \mathcal{V}T_0^{4}$. It is worth pointing out that electron-electron relaxation in metals is fast enough to secure a well defined electron temperature~\cite{Pothier1997}. 

For the ballistic channel discussed widely in the current manuscript, $G_{\rm th}\equiv G_{\rm Q}=\frac{\pi^2 k_{\rm B}^2}{3 h}T_0$, and we have for a general temperature difference exactly
\begin{equation} \label{M3}
	\dot Q = \frac{\pi^2 k_{\rm B}^2}{6 h}(T^2-T_0^2)= \frac{\pi^2 k_{\rm B}^2}{3 h}T_{\rm m}\delta T,
\end{equation} 
where $T_{\rm m}\equiv (T+T_0)/2$ is the mean temperature. 

In some experiments a differential two-absorber setup is preferable (see Fig.~\ref{epset}(b)). This allows for measurements of the temperatures of the two absorbers, $T_1$ and $T_2$, separately, and to determine the heat flux between the two without extra physical (wiring) connections for thermometry across the object of interest. In this case equations in this section apply if we replace $T,~T_0$ by $T_1,~T_2$, respectively. Such a setup offers more flexible calibration and sanity check options for the system, and also for tests of reciprocity (thermal rectification) by inverting the roles of source and drain, i.e., by reversing the temperature bias.

\subsection{Thermometry and temperature control}\label{Thermal Thermometry}
Here we comment very briefly on thermometry and temperature control in the experiments to be reported in this paper. The control of the local temperature is typically achieved by Joule heating applied to the electronic system. But depending on the type of the reservoir, this heat is either directly acting on the quantum conductor or indirectly, e.g. via the phonon bath. The simplest heating element is a resistive on-chip wire.
For heating (and local cooling), and in particular for thermometry, a hybrid normal metal-insulator-superconductor tunnel junction (NIS junction) is a common choice~\cite{Giazotto2006,Muhonen_2012,Courtois2014}. We defer discussion of this technique to Section~\ref{Experimental Heat}. In several experiments a simple resistive on-chip wire is used as a local heater. For thermometry one may use a similar wire and measure its thermal noise~\cite{Schwab2000}. Another option used in some recent experiments is to measure the current noise of a quantum point contact~\cite{Jezouin601,Banerjee2017}.

\section{Experimental setups, background information}\label{Experimental setups}
\subsection{Thermal conductance of a superconductor}\label{Experimental Thermal}
A superconductor obeying Bardeen-Cooper-Schrieffer theory (BCS theory)~\cite{Bardeen1957} forms an ideal building block for thermal experiments at low temperatures. A basic feature of a BCS-superconductor is its zero resistance, but in our context an even more important property is its essentially vanishing thermal conductance~\cite{Bardeen1959}. In bulk superconductors both electronic and the non-vanishing lattice thermal conductances play a role.

In small structures the exponentially vanishing thermal conductance at low temperatures can be exploited effectively to form thermal insulators that can at the same time provide perfect electrical contacts. In quantitative terms, according to the theory~\cite{Bardeen1959}, the ratio of the thermal conductivity $\kappa_{\rm e,S}$ in the superconducting state and $\kappa_{\rm e,N}$ in the normal state of the same material is given by
\begin{equation} \label{scth}
	\kappa_{\rm e,S}/\kappa_{\rm e,N}=\int_\Delta^\infty d\epsilon  \epsilon^2 f'(\epsilon) / \int_0^\infty d\epsilon  \epsilon^2 f'(\epsilon),
\end{equation}
where $\Delta \approx 1.76 k_{\rm B}T_{\rm C}$ is the gap of the superconductor with critical temperature $T_{\rm C}$. For temperatures well below $T_{\rm C}$, i.e., for $\Delta/k_{\rm B}T \gg 1$, we obtain an approximate answer for Eq. \eqref{scth} as  
\begin{equation} \label{scth2}
	\kappa_{\rm e,S}/\kappa_{\rm e,N}\approx \frac{6}{\pi^2} (\frac{\Delta}{k_{\rm B}T})^2 e^{-\Delta/k_{\rm B}T}.
\end{equation}
Since the normal state thermal and electrical conductivities are related by the Wiedemann-Franz law, we obtain
\begin{equation} \label{scth3}
	\kappa_{\rm e,S}\approx \frac{2\Delta^2}{e^2\rho T}  e^{-\Delta/k_{\rm B}T},
\end{equation}
where $\rho$ is the normal state resistivity of the conductor material. As usual, for the basic case of a uniform conductor with cross-sectional area $A$ and length $\ell$, we may then associate the thermal conductance $G_{\rm th}$ to thermal conductivity $\kappa$ as $G_{\rm th}=(A/\ell)\kappa$.

Aluminum and niobium are the most common superconductors used in the experiments described here. In many respects, Al follows BCS-theory accurately. In particular it has been shown~\cite{Saira2012} that the density of states (DoS) at energies inside the gap is suppressed at least by a factor $\sim 10^{-7}$ leading for instance to the exponentially high thermal insulation discussed here. The measured thermal conductivity of Al follows closely Eq.~\eqref{scth3}, shown in~\cite{Peltonen2010,Feshchenko2017}. At the same time Nb films suffer from non-vanishing subgap DoS, leading to power-law  thermal conductance in $T$, i.e. poor thermal insulation at the low temperature regime. In conclusion of this subsection we want to emphasize that Al is a perfect thermal insulator at $T \lesssim 0.3 T_{\rm C}$, except in the immediate contact with a normal metal leading to inverse proximity effect; this proximity induced thermal conductivity affects typically only within few hundred nm distances from a clean normal metal contact~\cite{Peltonen2010}. 

\subsection{Heat transport in tunneling}\label{Experimental Heat}
One central element in this review is a tunnel junction between two electrodes L (left) and R (right). The charge and heat currents through the junction can be obtained by perturbation theory where the coupling Hamiltonian between the electrodes is written as the tunnel Hamiltonian~\cite{Bruus2004} 
\begin{equation}\label{V}
	\hat{H}_c=\sum_{l,r}(t_{lr}\hat{a}_{l}^\dagger\hat{a}_{r}+t_{lr}^*\hat{a}_{l}\hat{a}_{r}^\dagger).
\end{equation}
Here $t_{lr}$ is the tunneling amplitude, and $\hat{a}_{l(r)}^\dagger$ and $\hat{a}_{l(r)}$ are the creation and annihilation operators for electrons in the left (right) electrode, respectively. 

To have the expression for number current from R to L one first obtains operator for it as $\dot{\hat{N}}_L=\frac{i}{\hbar}[\hat{H}_c,\hat{N}_L]$, where $\hat{N}_L=\sum_l\hat{a}_{l}^\dagger\hat{a}_{l}$ is the operator for the number of electrons in L. Then one can write the charge current operator as $\hat{I}=-e\dot{\hat{N}}_L$. 
In order to obtain the expectation value of the current that is measured in an experiment, $I\equiv\langle \hat{I}\rangle$, we employ linear response theory (Kubo formula~\cite{Kubo1957}) on the corresponding current operator, where $I=-i/\hbar\int_{-\infty}^{0}dt'\langle [\hat{I}(0),\hat{H}_c(t')]\rangle_0$, with $\langle .\rangle_0$ the expectation value in the unperturbed state. Assuming that the averages are given by the Fermi distributions in each lead, we have at voltage bias $V$
\begin{equation}\label{currtunn}
	I=\frac{1}{eR_T}\int d\epsilon~n_L(\tilde{\epsilon})n_R(\epsilon)[f_L(\tilde{\epsilon})-f_R(\epsilon)],
\end{equation}
where $\tilde{\epsilon}=\epsilon-eV$. Here the constant prefactor includes the inverse of the resistance $R_T$  of the junction such that $1/R_T=2\pi|t|^2\nu_L(0)\nu_R(0)e^2/\hbar$, with $|t|^2=|t_{rl}|^2={\rm constant}$, and $\nu_L(0),\nu_R(0)$ the density of states (DoS) in the normal state at Fermi energy in the left and right electrodes, respectively. Under the integral, $n_L(\epsilon),~n_R(\epsilon)$ are the normalized (by $\nu_L(0),~\nu_R(0)$, respectively) energy dependent DoSes, and $f_L(\epsilon),f_R(\epsilon)$ the corresponding energy distributions that are Fermi-Dirac distributions for equilibrium electrodes.

For heat current we use precisely the same procedure but now for the operator of energy of the left electrode $\hat{H}_{L}=\sum_{l}\epsilon_{l}\hat{a}_{l}^\dagger\hat{a}_{l}$, instead of the number operator, where $\epsilon_l$ is the energy of a single particle state in L. Then we find the expectation value of the heat current out from L electrode, $\dot Q_L =-\langle \dot{H}_L\rangle$ as
\begin{equation}\label{heattunn}
	\dot{Q}_L=\frac{1}{e^2R_T}\int d\epsilon~ \tilde{\epsilon}~n_L(\tilde{\epsilon})n_R(\epsilon)[f_L(\tilde{\epsilon})-f_R(\epsilon)].
\end{equation} 
Here we may briefly comment on the relation of energy and heat currents, $\mathcal{J}$ and $\dot{Q}$ introduced in Section~\ref{Thermoelectric transport}. Inserting $\tilde{\epsilon}=\epsilon-eV$, we find immediately $\dot{Q}_L=\mathcal{J}-IV$, where $\mathcal{J}\equiv{(e^2R_T)}^{-1}\int d\epsilon~ \epsilon~n_L(\tilde{\epsilon})n_R(\epsilon)[f_L(\tilde{\epsilon})-f_R(\epsilon)]$. Writing the equation for the heat from the right electrode in analogy to Eq.~\eqref{heattunn}, we find $\dot{Q}_R=-\mathcal{J}$. Thus we have $\dot{Q}_L+\dot{Q}_R=-IV$, which presents energy conservation: the total power taken from the source goes into heating the two electrodes. This is natural since in steady state work equals heat, as the internal energy of the system is constant.
\begin{figure}[h!]
	\centering
	\includegraphics [width=\columnwidth] {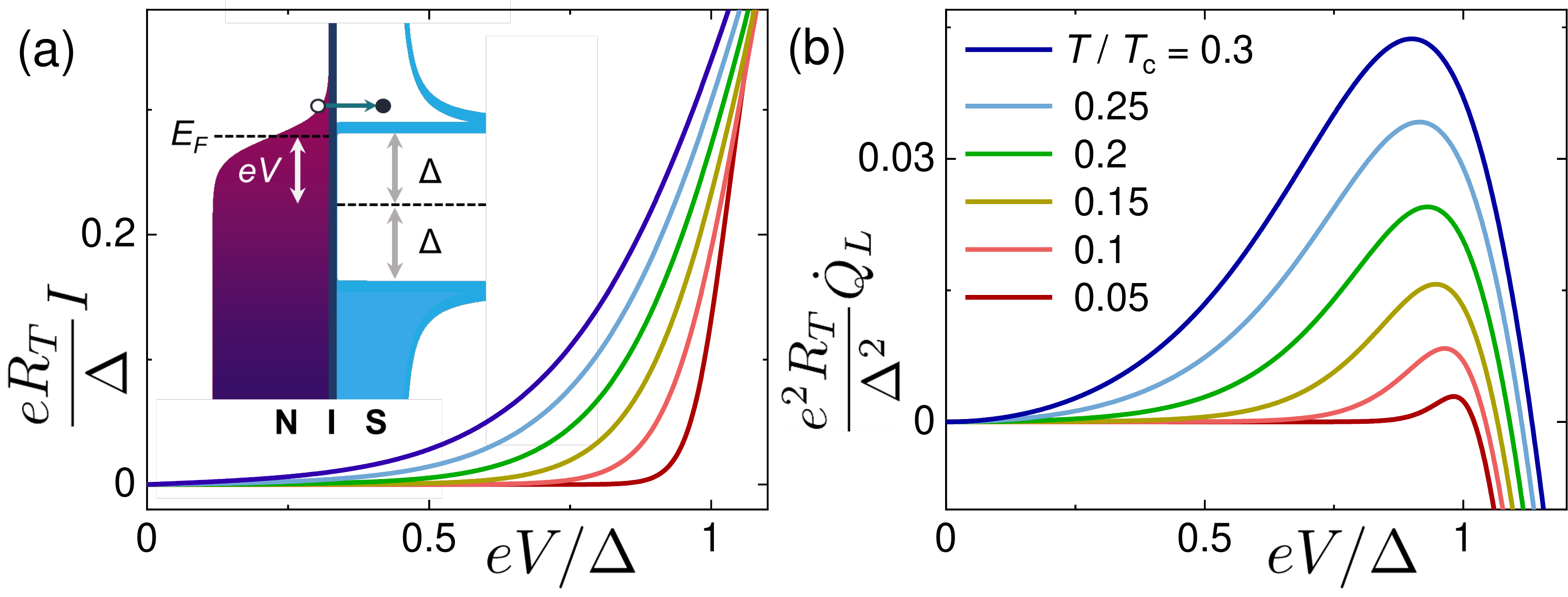}
	\caption{Properties of an NIS tunnel junction. The inset in (a) shows the energy diagram of a biased by voltage $V$ junction between a normal metal (N) and superconducting (S) electrode connected via an insulating (I) barrier. Due to the BCS gap $\Delta$ in S, transport is blocked at $eV\ll \Delta$. At a voltage close to the gap value, as in the figure, electrons at the highest energy levels can tunnel to the superconductor as shown, leading to both non-vanishing charge current and cooling of N. The main frame of (a) shows calculated current-voltage curves at different values of $T/T_C=0.05-0.3$ from bottom to top (for both panels): at these subgap voltages the junction provides a sensitive thermometer. (b) presents similarly calculated power $\dot Q_L$ versus $V$ curves, demonstrating cooling of N at $eV \lesssim \Delta$. At higher voltages $eV \gg \Delta$, $\dot Q_L$ becomes negative, meaning it serves as a Joule heater of N.
		\label{NIS-IV-Cooling}}
\end{figure}
\begin{figure}[h!]
	\centering
	\includegraphics [width=0.35\textwidth] {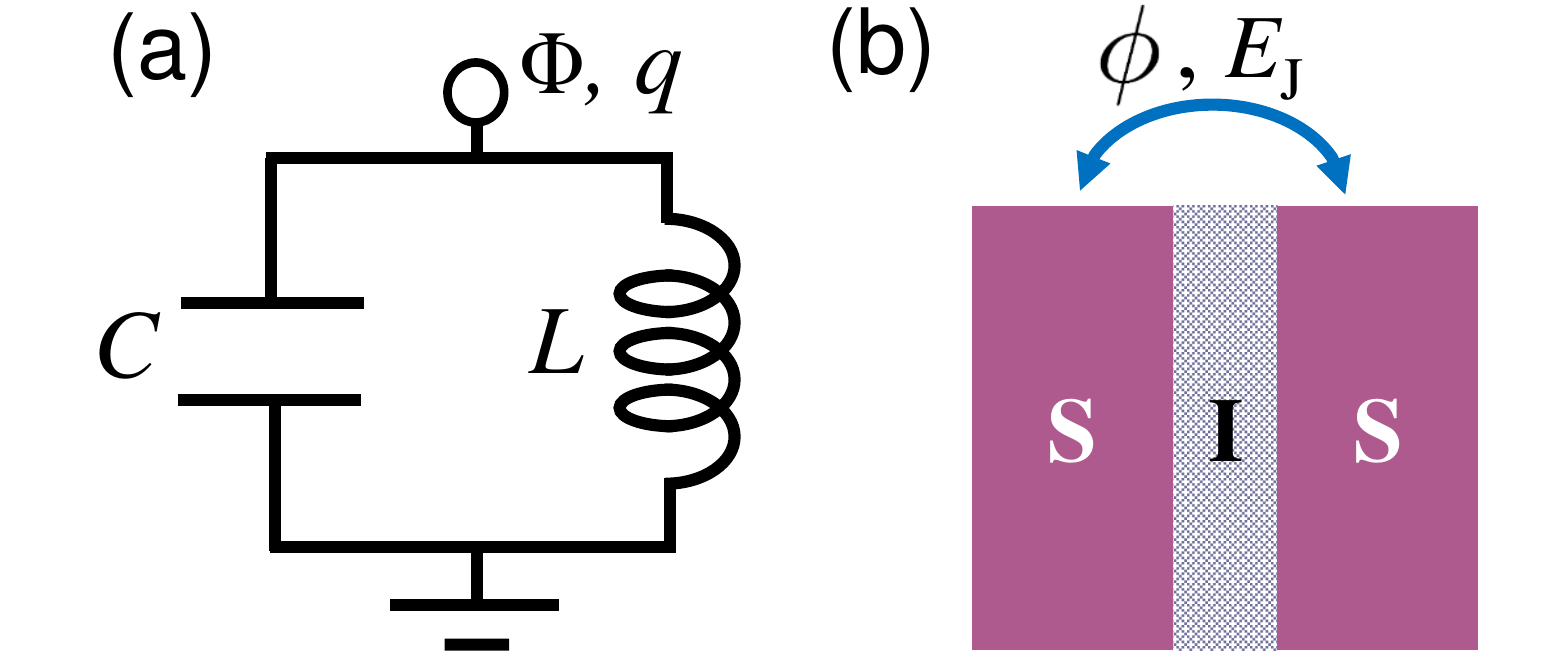}
	\caption{Central elements of superconducting quantum devices. (a) $LC$ circuit with flux $\Phi$ and charge $q$. (b) Josephson junction with phase difference $\phi$ and Josephson energy $E_{\rm J}$.
		\label{LCSIS}}
\end{figure}

As the most basic example of both the electrodes being normal metal (NIN junction, I for insulator), we have $n_L(\epsilon)=n_R(\epsilon)=1$.  Equations~\eqref{currtunn} and \eqref{heattunn} then yield under quite relaxed conditions $I=V/R_T$ and  $\dot{Q}_L=-\frac{V^2}{2R_T}$, i.e. the junction is ohmic and the Joule power is dissipated equally to the two electrodes.

Another important example is a normal-superconductor junction (NIS junction, L=N, R=S), Fig.~\ref{NIS-IV-Cooling}. Its usefulness in thermometry (see Fig.~\ref{NIS-IV-Cooling}(a)) is based on the superconducting gap $\Delta$ that leads to non-linear, temperature dependent current-voltage characteristics. This feature probes the temperature of the normal side of the contact. Such a temperature dependence is universal, $d\ln (I/I_0)/dV = e/(k_{\rm B}T)$, where $I_0=\sqrt{\pi \Delta k_{\rm B}T/2}/(eR_T)$, making NIS a primary thermometer in principle. This is strictly speaking true only for an ideal junction with low transparency. Therefore the common practice is to use it as a secondary thermometer~\cite{Lounasmaa1974}, meaning that one measures a thermometric response of it near equilibrium, for instance the voltage at a small fixed current, against the independently measured temperature of the cryostat (heat bath). The other important feature of the NIS junction lies in its thermal properties. When biased at a voltage of about $\Delta/e$, heat is carried away from the N side (and S is heated), i.e., it acts as a refrigerator (see Fig.~\ref{NIS-IV-Cooling}(b)). At $V \gg \Delta/e$ the junction provides usual Joule heating. This is how a NIS junction can be used both as a cooler and heater of a mesoscopic reservoir. Numerically calculated current-voltage and cooling power characteristics, together with a schematic energy diagram have been depicted in the figure. The main characteristics of a NIS junction, based on analytical approximations at low temperatures  are: $I\approx I_0 e^{-\Delta/k_{\rm B}T}$ at voltages below the gap, and the maximal cooling of normal metal at $eV\approx \Delta$ is $\dot Q_L^{\rm max}\approx + 0.59 (\Delta^2/e^2R_T)(k_{\rm B}T/\Delta)^{3/2}$. 

Microrefrigeration by electron transport is a technique that has been reviewed elsewhere~\cite{Courtois2014,Giazotto2006,Muhonen_2012}. For an interested reader, the following non-exhaustive list of original references on the topic besides the mentioned review articles could be recommended:~\cite{Prance2009,M.Nahum1994,Leivo1996,Kuzmin_2004,Clark2004,Nguyen_2013,Feshchenko2014}.

\subsection{Hamiltonian of a quantum circuit}\label{Experimental Hamiltonian}
Another key element in our context is a harmonic oscillator, and in some cases a non-linear quantum oscillator, usually in form of a Josephson junction~\cite{Tinkham2004}. To avoid dissipation the linear harmonic oscillator in a circuit is commonly made of a superconductor, often in form of a coplanar wave resonator~\cite{Krantz2019}. The Hamiltonian of such an $LC$-oscillator, shown in Fig.~\ref{LCSIS}(a) is composed of the kinetic $q^2/(2C)$ and potential $\Phi^2/(2L)$ energies, respectively, where $q$ is the charge on the capacitor and $\Phi$ the flux of the inductor.

The charge is the conjugate momentum to flux as
$q=C\dot{\Phi}$, and the total Hamiltonian is then 
\begin{eqnarray}\label{HO}
	\hat{H}=\frac{\hat{q}^2}{2C}+\frac{\hat{\Phi}^2}{2L},
\end{eqnarray} 
i.e. that of a harmonic oscillator, with $\hat{q}$ and $\hat{\Phi}$ as charge and flux operators, respectively. Introducing the creation $\hat{c}^\dagger$ and annihilation $\hat{c}$ operators such that $[\hat{c},\hat{c}^\dagger]=1$,
we have
\begin{equation}\label{PhiQ}
	\hat{\Phi}=\sqrt{\frac{\hbar Z_0}{2}}(\hat{c}+\hat{c}^\dagger),\,\, \hat{q}=-i\sqrt{\frac{\hbar}{2Z_0}}(\hat{c}-\hat{c}^\dagger),
\end{equation}
yielding the standard harmonic oscillator Hamiltonian
\begin{eqnarray}\label{H3}
	H=\hbar\omega_0(\hat{c}^\dagger \hat{c}+\frac{1}{2}),
\end{eqnarray}
where $\omega_0=1/\sqrt{LC}$, and $Z_0=\sqrt{L/C}$ are the (angular) frequency and impedance of the oscillator.

For a Josephson tunnel junction, shown in Fig.~\ref{LCSIS}(b), the Josephson relations~\cite{Josephson1962} are
\begin{equation}\label{JJ}
	\hbar\dot{\phi}=2eV, \,\, I=I_c\sin\phi,
\end{equation}
where $\phi$ is the phase difference across the junction related to flux by $\phi=(2e/\hbar)\Phi$. In the second Josephson relation, $I$ is the current through the junction. The sinusoidal current-phase relation applies strictly for a tunnel junction, with critical current $I_c$. For different types of weak links, sinusoidal dependence does not necessarily hold~\cite{Tinkham2004}. The energy stored in the junction (=work done by the source) is then obtained for a current biased case from $I=\partial E/\partial \Phi$ as
\begin{eqnarray}\label{energy2}
	E=\int^{\Phi}I~d\Phi= -E_J\cos \phi.
\end{eqnarray}
This constitutes the Josephson part of the Hamiltonian, also called $\hat{H}_J$. For small values of $\phi$, ignoring the constant part, we have
\begin{eqnarray}\label{energy3}
	E\simeq  \frac{\Phi^2}{2L_J},
\end{eqnarray}
where $L_J=\hbar/(2eI_c)$ is the Josephson inductance. Therefore, in the “linear regime” a Josephson junction can be considered as a harmonic oscillator such that Eqs.~\eqref{HO}-\eqref{H3} apply with $L$ replaced by $L_J$. Yet the actual non-linearity of a Josephson junction makes it an invaluable component in quantum information processing and in quantum thermodynamics. 

A magnetic flux tunable Josephson junction, for instance in form of two parallel junctions with a superconducting loop in between, is the renowned superconducting quantum interference device (SQUID) to be discussed in later sections.
\begin{figure}
	\centering
	\includegraphics [width=\columnwidth] {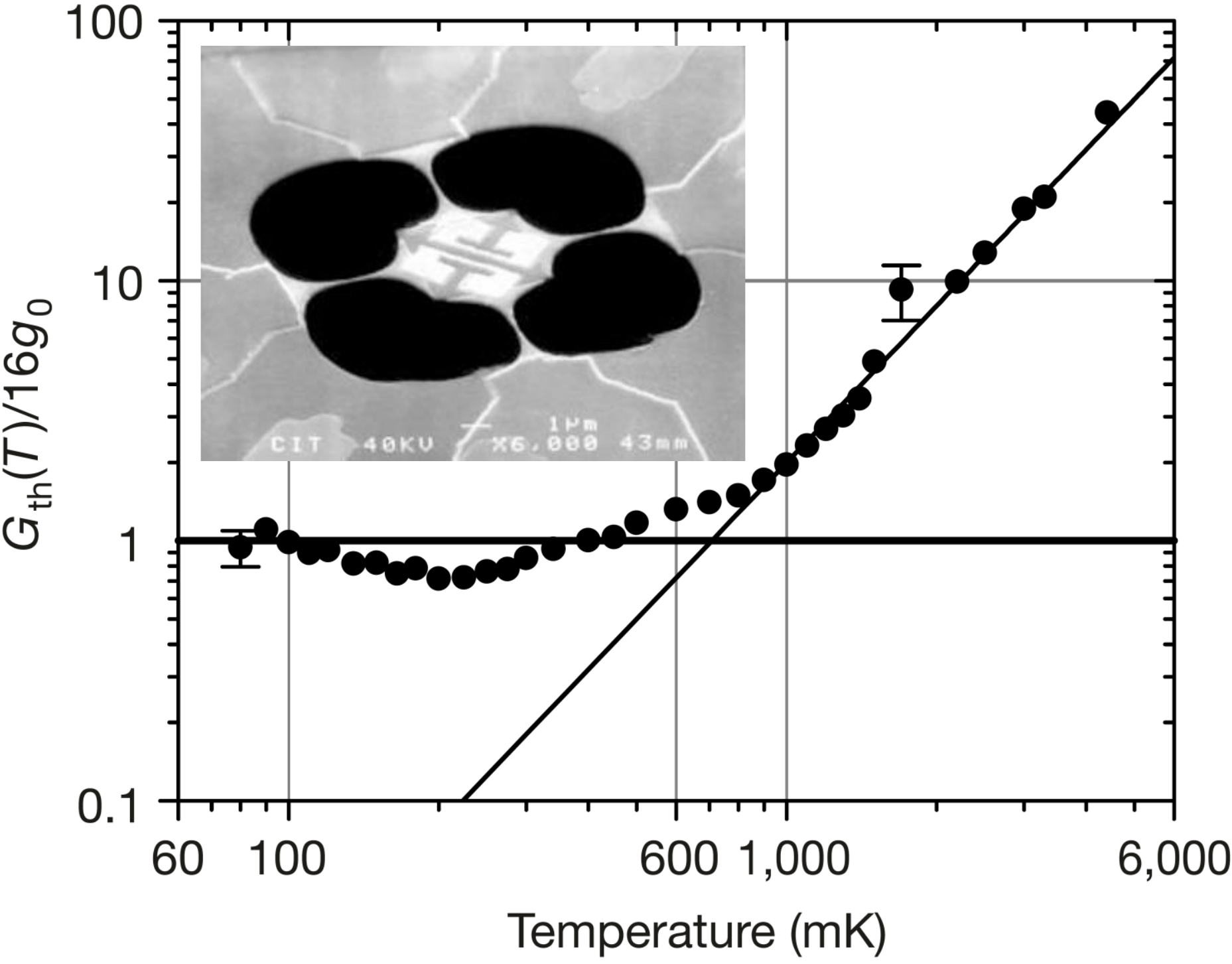}
	\caption{View of the suspended structure of~\cite{Schwab2000} for measuring quantized thermal conductance. (Inset) In the center, $4\times 4~\mu{\rm m}^2$ phonon cavity is patterned from the membrane; the bright areas on the central membrane are Au-thin-film transducers connected to Nb-thin-film leads on top of phonon waveguides. The membrane has been completely removed in the dark regions. (main panel) Temperature dependence of the measured thermal conductance normalized by $16 G_{\rm Q}~(16g_0)$. Adapted from~\cite{Schwab2000}.
		\label{phonon-roukes}}
\end{figure}

\subsection{Quantum noise of a resistor}\label{Experimental Quantum}
The quantum noise of a resistor is an important quantity as it determines the heat emission and absorption in form of thermal excitations. In later sections it becomes obvious how this noise yields the Joule power in a circuit. 

Consider that the resistor in the quantum circuit is formed of a collection of harmonic oscillators with ladder operators $\hat{b}_i$ and $\hat{b}_i^\dagger$ with frequencies $\omega_i$. The phase operator in the interaction picture reads
\begin{equation}\label{phase2}
	\phi(t)=\sum_{i} \lambda_i (\hat{b}_ie^{-i\omega_it}+\hat{b}_i^\dagger e^{i\omega_it}),
\end{equation}
with coefficients $\lambda_i$. The voltage fluctuations are related to phase as $v(t)=\frac{\hbar}{e}\dot{\phi}(t)$
\begin{eqnarray}\label{voltage1}
	v(t)=i\frac{\hbar}{e}\sum_{i} \lambda_i\omega_i(\hat{b}_i^\dagger e^{i\omega_it}-\hat{b}_ie^{-i\omega_it}).
\end{eqnarray}
Then the spectral density of voltage noise $S_v(\omega)=\int_{-\infty}^{\infty} dt e^{i\omega t}\langle v(t)v(0)\rangle$ is given by
\begin{eqnarray}\label{Sv4}
	S_v(\omega)=\frac{2\pi\hbar^2}{e^2}\int_{0}^{\infty} d\Omega~\nu(\Omega)~\lambda(\Omega)^2~&&\Omega^2\bigg \{[1+n(\Omega)]\delta(\omega-\Omega)\nonumber\\&&+n(\Omega)\delta(\omega+\Omega)\bigg \},
\end{eqnarray}
where $\nu(\Omega)$ is the oscillator density of states. Now we consider both positive and negative frequencies, which correspond to quantum emission and absorption processes. For positive frequencies only the first term survives as
\begin{equation} \label{posfreq}
	S_v(\omega)=\frac{2\pi\hbar^2}{e^2}\nu(\omega)\lambda(\omega)\omega^2[1+n(\omega)]
\end{equation}
Considering similarly the negative frequencies, we find
\begin{equation}\label{db}
	S_v(-\omega)=e^{-\beta \hbar\omega}S_v(\omega),
\end{equation}
which is the famous {\sl detailed balance condition}.

We know that the classical Johnson-Nyquist noise~\cite{Johnson1928,Nyquist1928} of a resistor at $k_{\rm B}T\gg \hbar\omega$ reads
\begin{eqnarray}\label{classic}
	S_v(\omega)=2k_{\rm B}T R.
\end{eqnarray}
This is the classical fluctuation-dissipation theorem (FDT~\cite{Callen1951}) applied to the resistor.
In this limit by using Taylor expansion, we have $(1-e^{-\beta\hbar\omega})^{-1}\simeq(\beta\hbar\omega)^{-1}$, so that using Eq.~\eqref{posfreq} we have connection between the oscillator properties and the physical resistance as~\cite{Karimi2021}
\begin{eqnarray}\label{resistor}
	\lambda_i^2=\frac{Re^2}{\pi\hbar\nu(\omega_i)\omega_i}.
\end{eqnarray}
Substituting this result in Eq.~\eqref{Sv4}, we obtain at all frequencies
\begin{eqnarray}\label{positive2}
	S_v(\omega)=2R\frac{\hbar\omega}{1-e^{-\beta\hbar\omega}}.
\end{eqnarray}

\begin{figure}
	\centering
	\includegraphics [width=\columnwidth] {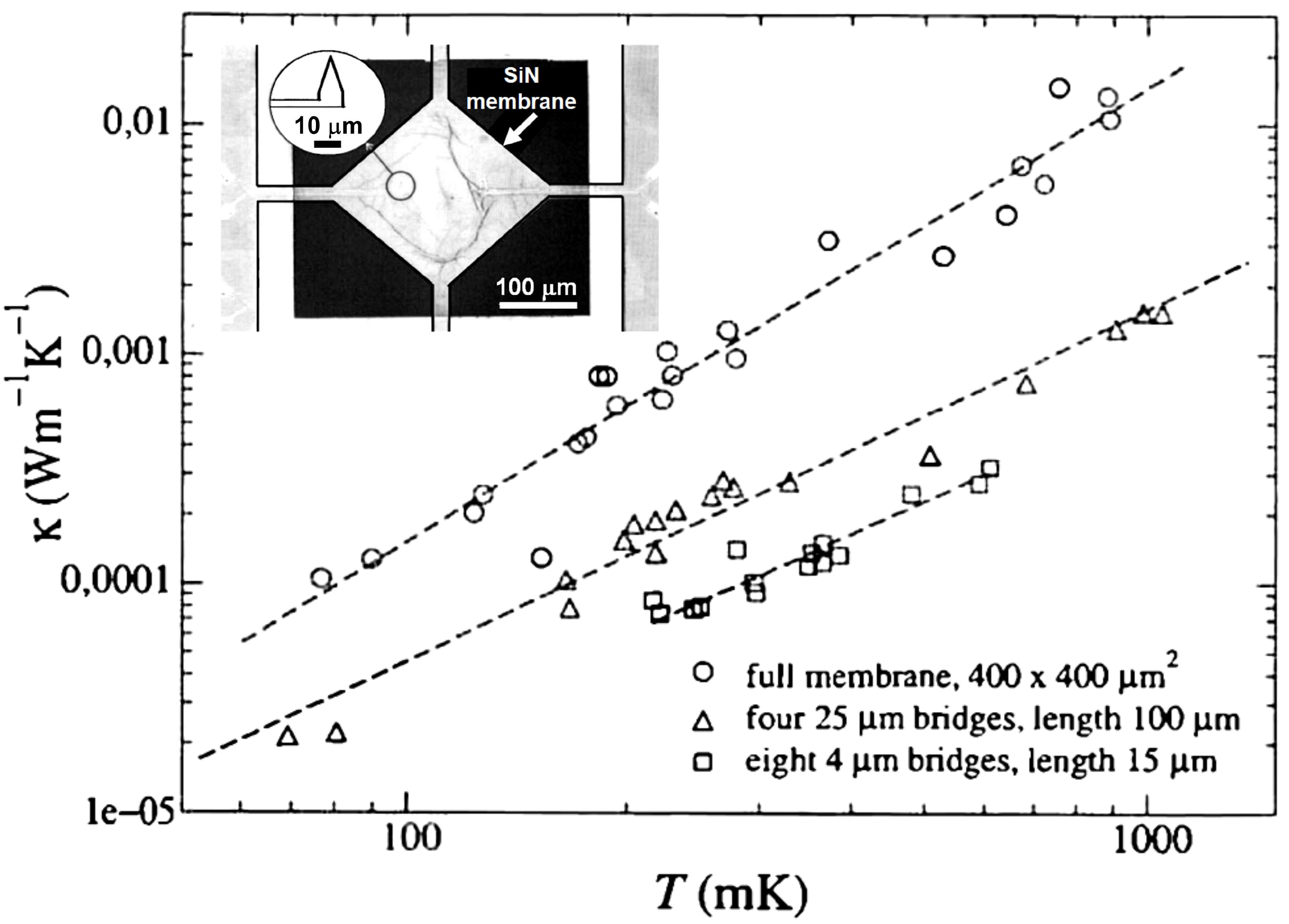}
	\caption{Thermal conductivity $\kappa$ of a $200$~nm-thick silicon nitride membrane measured in three different geometries as a function of membrane temperature. Dashed lines present the fitted functions: $\kappa\simeq 14.5T^{1.98}$~mWm$^{-1}$K$^{-1}$ for full membrane, $\kappa\simeq 1.58T^{1.54}$~mWm$^{-1}$K$^{-1}$ and $\kappa\simeq 0.57T^{1.37}$~mWm$^{-1}$K$^{-1}$ for $25~\mu$m and $4~\mu$m wide bridges, respectively, where $T$ is expressed in K. The data of $400\times 400~\mu$m$^2$ full membrane and $25~\mu$m-wide bridge are presented in~\cite{Leivo1998} while the one for $4~\mu$m-wide bridge is unpublished~\cite{Leivothesis}. The corresponding thermal conductance $G_{\rm th}=\kappa A/L$ for one bridge with area $A$ and length $L$ at $T=0.1$ K for both $25~\mu$m and $4~\mu$m are $2.3\times 10^{-12}$~$\rm W/\rm K$ and $1.3\times 10^{-12}$~$\rm W/\rm K$ that give $N\simeq 24$ and $N\simeq 14$, respectively, assuming fully ballistic channels. Adapted from~\cite{Leivo1998,Leivothesis}.
		\label{leivo}}
\end{figure}

\section{Phonons}\label{Phonons}
Quantized thermal conductance was demonstrated experimentally for the first time by~\cite{Schwab2000}. In their setup, as shown in the inset of Fig.~\ref{phonon-roukes}, the "phonon cavity" consists of a $4\times 4~\mu{\rm m}^2$ block of silicon nitride membrane with $60$~nm thickness suspended by four legs of equal thickness. Each leg has catenoidal waveguide shape whose diameter at the narrowest point is less than $200$~nm. This waveguide shape as a 1D channel is the ideal profile to achieve unit transmissivity between the suspended cavity and the bulk reservoir~\cite{Rego1998}. Two Au-film resistors with $25$~nm thickness were patterned on the suspended central block; one of them serves to apply the Joule heating to generate the temperature gradient along the legs and the other one worked as a thermometer to measure the phonon cavity temperature. The electron temperature of the resistor was measured by a low noise amplifier (dc SQUID) operating with nearly quantum-limited energy sensitivity by measuring the electrical Johnson noise of the resistor.

The measurement of~\cite{Schwab2000} probes the thermal conductance by phonons across the four silicon nitride bridges as a function of bath temperature. These data are shown in the main frame of Fig.~\ref{phonon-roukes}. The result exhibits the usual phononic thermal conductance, $\propto T^3$, at temperatures above 1~K. Below this temperature there is a rather abrupt leveling off of $G_{\rm th}$ to the value $16 G_{\rm Q}$ (here the notation is such that $g_0 \equiv G_{\rm Q}$). The authors argue that the coefficient $16$ arises from the trivial factor $4$ due to four independent bridges in the structure and the less trivial factor $4$ due to four possible acoustic vibration modes of each leg in the low temperature limit: one longitudinal, one torsional, and two transverse modes. In later theoretical works the somewhat meandering behavior of $G_{\rm th}/G_{\rm Q}$ below the crossover temperature was explained to arise from remaining scattering of phonons in the bridges, i.e. from non-ballistic transport, whose effect is expected to get weaker in the low temperature limit~\cite{Santamore2001}.

Over the years, there have been a few other experiments on thermal conductance by phonons in restricted geometries. The one by Leivo~\cite{Leivo1998,Leivothesis,Manninen1997} employed 200~nm-thick silicon nitride membranes in various geometries (see Fig.~\ref{leivo}). The experiments were performed by applying Joule heating on a central membrane in an analogous way as in the experiment of~\cite{Schwab2000}, and the resulting temperature change to obtain the thermal conductance was then read out by measuring the temperature dependent conductance of NIS probes processed on top of the same membrane. In this case the wiring running along the bridges was made of aluminum which is known to provide close to perfect thermal isolation at temperatures well below the superconducting transition at $T_{\rm C}\approx 1.4$~K, see Section~\ref{Experimental Thermal}. In general, there are many conduction channels in the wide bridges, as demonstrated in the figure caption. Yet this number for a single $w=4$~$\mu$m wide bridge is $N=14$ at $T=100$~mK, which is already in the same ballpark as the prediction of $N=4$ of~\cite{Rego1998}. Naturally the ballisticity of these 15~$\mu$m long bridges is unknown though. Yet these experiments provide evidence of thermal conductance close to the quantum limit.

The experiment of~\cite{Schwab2000} was followed by several measurements in different temperature ranges and materials. Experiments on GaAs phonon bridges of sub-$\mu$m lateral dimensions were earlier performed at temperatures above 1~K~\cite{Tighe1997} and later down to 25~mK bath temperature~\cite{C.S.Yung2002}. The latter experiment measuring the temperature of the GaAs platform in the middle using NIS tunnel junctions demonstrated Debye thermal conductance at $T \gg 100$~mK, but tended to follow the expected quantum thermal conductance at the lowest temperatures. In the more recent experiments by~\cite{AdibTavakoli2018,Tavakoli2017} the measurement on sub-micron wide silicon nitride bridges was made differential in the sense that there was no need to add superconducting leads on these phonon-conducting legs. The results at the lowest temperatures of $\sim 0.1$~K fall about one order of magnitude below the quantum value, and the temperature dependence of thermal conductance is close to $T^2$. The authors of~\cite{AdibTavakoli2018,Tavakoli2017} propose non-ballistic transmission in their bridges to be the origin of their results. Finally experiments by~\cite{Zen2014} demonstrate that thermal conductance can be suppressed strongly even in two-dimensions by proper patterning of the membranes into a nanostructured periodic phononic crystal. 
\begin{figure}
	\centering
	\includegraphics [width=\columnwidth] {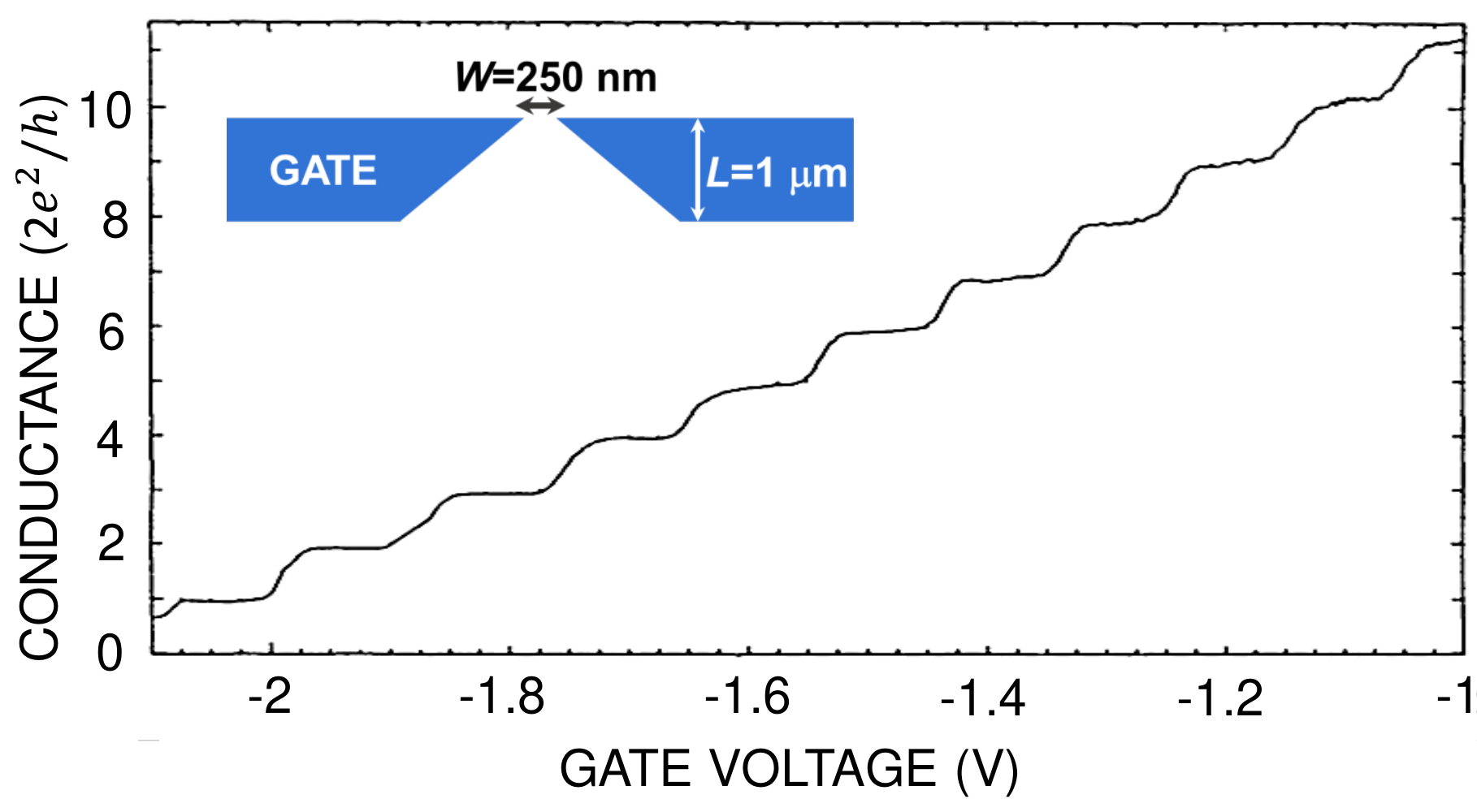}
	\caption{Measured quantized conductance of a point contact in a two-dimensional electron gas as a function of gate voltage. The conductance demonstrates plateaus at multiples of $2e^2/h$. The layout of the point contact is shown schematically in the inset. Adapted from~\cite{Wees1988}.
		\label{vanWees}}
\end{figure}

\section{Electrons and fractional charges}\label{Electrons}
Charged particles play a special role in assessing quantum transport properties, since they provide a straightforward access to both particle number current and heat current. For instance, in the case of electrons, we can count the carriers by measuring directly the charge current and the associated conductance. When the mean free path of the carriers is much larger than the physical dimensions of the contact, transport can become ballistic.  According to Eq.~\eqref{G1}, the electrical conductance then assumes only integer multiple values of elementary conductance quantum. The very first experiments on quantized conductance of a point contact in a GaAs-AlGaAs two-dimensional high mobility electron gas (2DEG) hetrostructures were performed by~\cite{Wees1988,Wharam1988}. In~\cite{Wees1988} the point contact was formed by a top metallic gate with a width $W\simeq 250$~nm opening in a tapered geometry to form a voltage-controlled narrow and short channel in the underlying electron gas. The layout of the gate electrode is shown in the inset of Fig.~\ref{vanWees}. At negative gate voltages electrons are repelled under the gate and the width of the channel for carriers is $\lesssim 100$~nm which is well below the mean free path of $l\simeq 8.5~\mu$m. The measured conductance of the point contact shown in Fig.~\ref{vanWees} exhibits well defined plateaus at the expected positions $N2e^2/h$ as a function of applied gate voltage~\cite{Wees1988}. The factor 2 with respect to Eq.~\eqref{G1} arises from spin degeneracy. 
\begin{figure}
	\centering
	\includegraphics [width=\columnwidth] {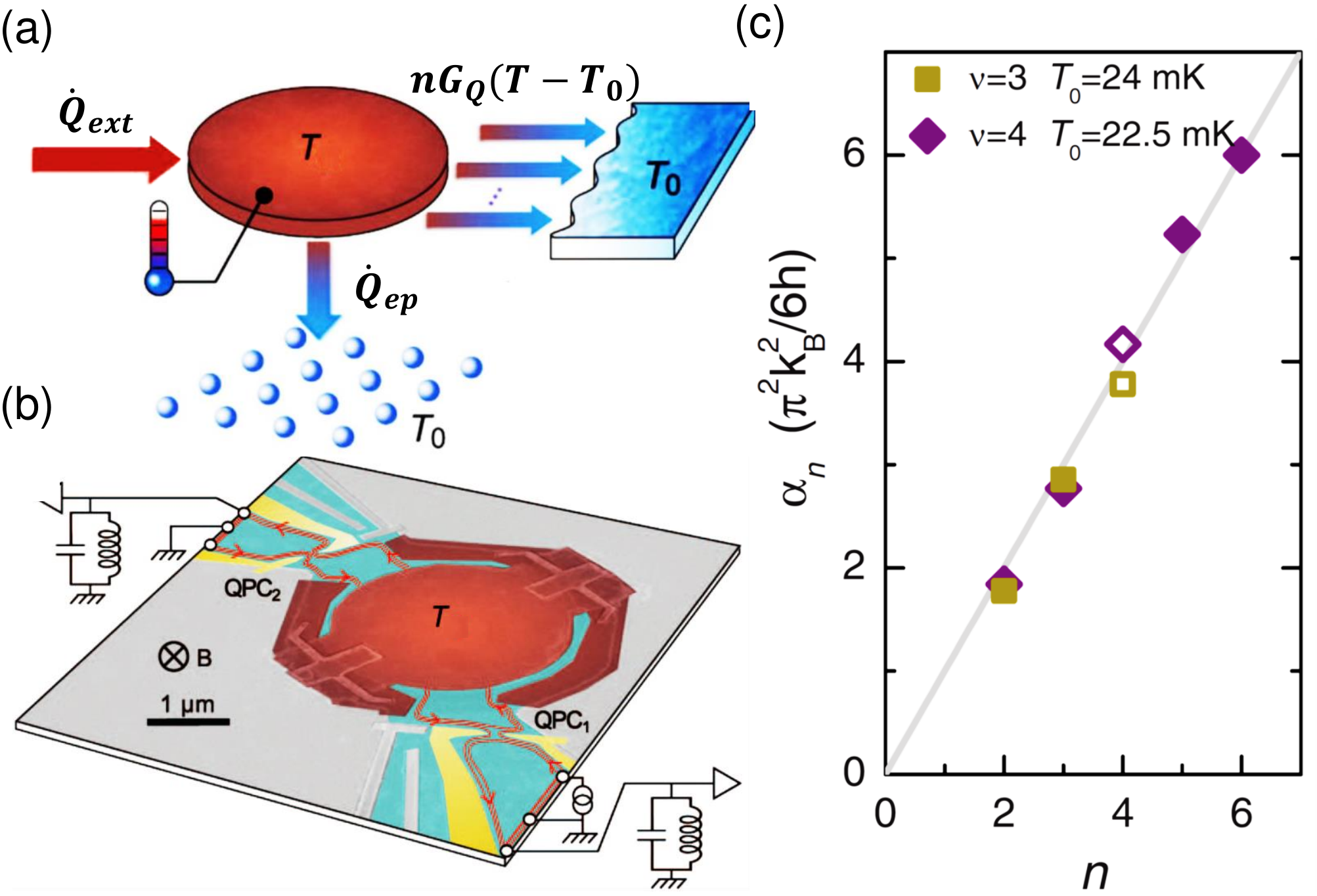}
	\caption{Measuring quantized heat carried by electrons. (a) Applying Joule power $\dot{Q}_{\rm ext}$ to a metal plate (brown disk), the electronic temperature increases up to $T$ and then the heat flows via $n$ ballistic quantum channels to reservoir and to the phonon heat bath $\dot{Q}_{\rm ep}$, which both have fixed temperature $T_0$. (b) The colored scanning electron micrograph of the measured sample. In the center, the metallic ohmic contact in brown is connected to two quantum point contacts ($\rm QPC_1$ and $\rm QPC_2$) in yellow (lightest area) via a two-dimensional Ga(Al)As electron gas in light green (surrounding the point contacts). The red lines with arrows around the metal plate indicate the two propagating edge channels ($\nu=3$ or $\nu=4$). The Joule power is applied to the metallic plate through a QPC, and the two $LC$-tank circuits are for noise thermometry measurement. (c) The grey line shows the predictions for quantum limit of heat flow while the symbols exhibit the extracted electronic heat current normalized by $\pi^2k_{\rm B}^2/6h$ as a function of number of electronic channels $n$. Adapted from~\cite{Jezouin601}.
		\label{Jezouin-Qthercondu}}
\end{figure}

Thirty years after the experiments on quantized electrical conductance by electrons~\cite{Wees1988,Wharam1988}, Jezouin et al.~\cite{Jezouin601} measured the quantum-limited heat conductance of electrons in a quantum point contact. The principle and practical implementation of this experiment and its setup are shown in Fig.~\ref{Jezouin-Qthercondu}(a) and \ref{Jezouin-Qthercondu}(b). A micrometer-sized metal plate is connected to both cold phonon bath and to a big electronic reservoir via an adjustable number $n$ of ballistic quantum channels with both reservoirs at $T_0$ as shown in Fig.~\ref{Jezouin-Qthercondu}(a). By injecting Joule power $\dot{Q}_{\rm ext}$ to the metallic plate, the electrons were heated up to temperature $T$ which can be directly measured by a noise thermometer. This power is then transmitted via the $n$ quantum channels at the rate $nG_{\rm Q}(T-T_0)$ through two quantum point contacts ($\rm QPC_1$ and $\rm QPC_2$) and to the phonon bath at rate $\dot{Q}_{\rm ep}$ which is independent of $n$. Both QPCs display clear plateaus of the measured electrical conductance at $n_1e^2/h$ and $n_2e^2/h$, respectively, where $n_1$ and $n_2$ are integers. The sum $n=n_1+n_2$ determines the number of quanta carrying the heat out of the plate electronically. The structure used in this experiment~\cite{Jezouin601} satisfies the conditions of having sufficient electrical and thermal contact between the metal plate and the two-dimensional electron gas underneath. Moreover the thermal coupling to the phonon bath and via the QPCs is weak enough such that the central electronic system forms a uniform Fermi gas (fast electron-electron relaxation and diffusion across the plate) at temperature $T$. A perpendicular magnetic field was applied to the sample for it to be in the integer quantum Hall effect regime at filling factors $\nu=3$ or $\nu=4$. Figure~\ref{Jezouin-Qthercondu}(c) shows $\alpha_n$, the measured electronic heat conductance normalized by $\pi^2k_{\rm B}^2/6h$ as a function of the number $n$ of electronic channels as symbols which fall on a straight line with unit slope shown by the grey line, thus demonstrating the quantized thermal conductance at the expected level. Equivalently this experiment demonstrates Wiedemann-Franz law on the current plateaus.

The work of~\cite{Jezouin601} was preceeded already by two experiments, some two decades earlier~\cite{Molenkamp1992,Chiatti2006}, where $G_{\rm Q}$ was tested with an order of magnitude accuracy. Both these measurements were performed on GaAs-based 2DEGs, and in both of them, thermal conductance was obtained by measuring the Seebeck coefficient (thermopower) and extracting the corresponding temperature difference. In~\cite{Molenkamp1992}, they then determined $G_{\rm th}$ which is within a factor two agreement with the assumption that Wiedemann-Franz law applies on their conduction plateaus of the QPC. In~\cite{Chiatti2006}, they made a similar experiment with the same philosophy but with improved control of the structure and system parameters. With these assumptions they come to a good agreement between thermal conductance and electrical conductance via the Wiedemann-Franz law.

\begin{figure}
	\centering
	\includegraphics [width=\columnwidth] {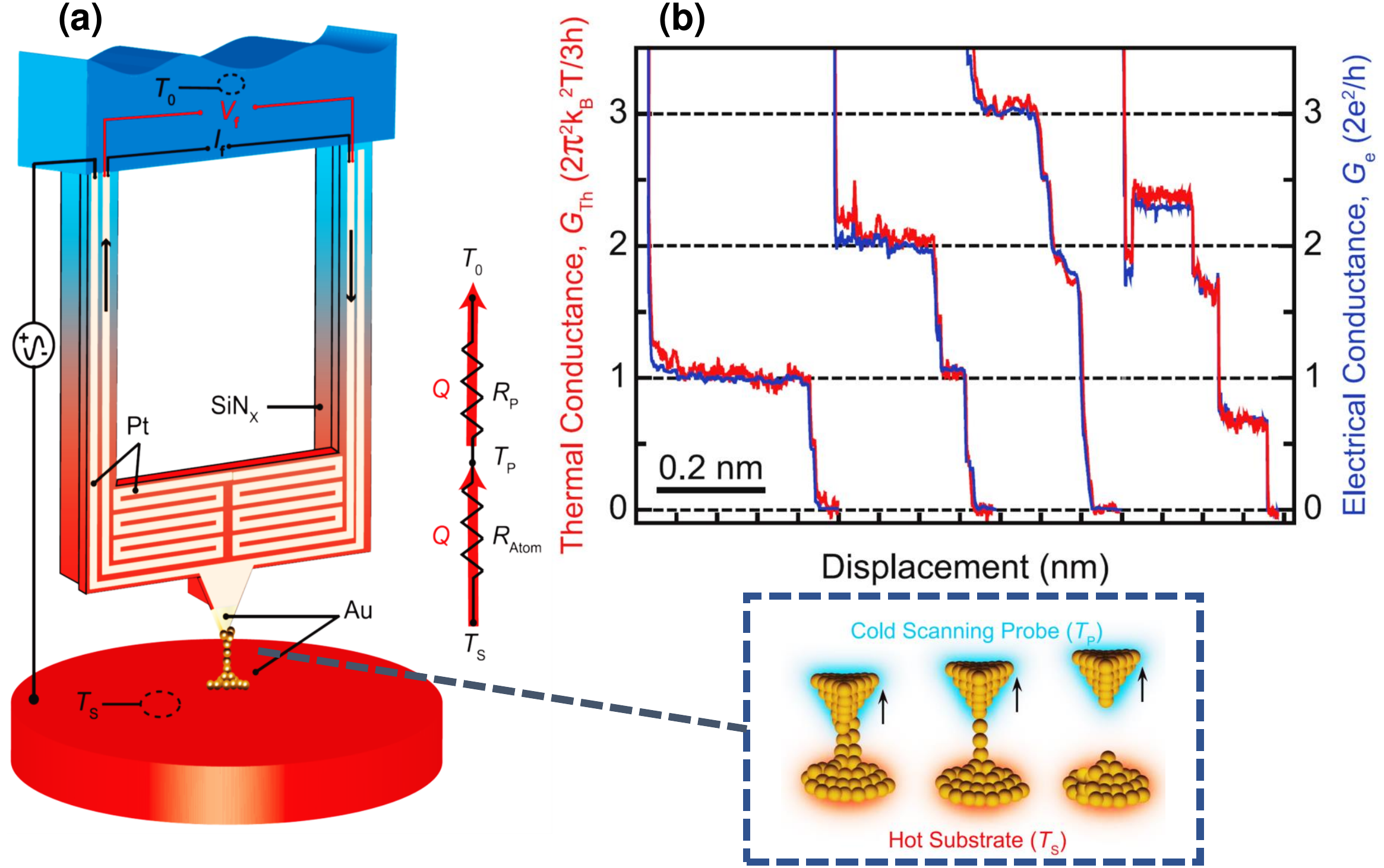}
	\caption{Experimental setup and results on quantized thermal conductance in single atom junctions. (a) A calorimetric scanning thermal microscopy probe, which shows schematically how to make atomic junctions connected to a heated metallic substrate. By applying a small voltage bias and measuring the resulting current, the electrical conductance of the tip-substrate junction can be measured. $T_{\rm S}$ and $T_0$ are the temperatures of the substrate and the thermal reservoir, respectively. The zoom-out describes schematically the atomic chains forming, narrowing, and breaking during the withdrawal of the probe from the heated substrate. (b) The almost overlapping measured thermal (red, left) and electrical (blue, right) conductance traces normalized by $2\pi^2k_{\rm B}^2T/3h$ and $2e^2/h$, respectively. Adapted from~\cite{Cui1192}.
		\label{CuiRT}}
\end{figure}

In recent years, it has become possible to measure quantized thermal conductance even at room temperature~\cite{Cui1192,Mosso2017}. The experiments are performed on metallic contacts of atomic size with scanning thermal microscopy probes. The material of choice is typically Au, although experiments on Pt have also been reported~\cite{Cui1192}. The setup and experimental observations of~\cite{Cui1192} are presented in Fig.~\ref{CuiRT}. The electrical conductance plateaus at multiples of $2e^2/h$ are typically seen when pulling the contact to the few conductance channel limit. The remarkable feature in the data is that the simultaneous thermometric measurement confirms the Wiedemann-Franz law for electric transport within 5 - 10\% accuracy, thereby demonstrating quantized thermal conductance~\cite{Cui1192}. 

\begin{figure}
	\centering
	\includegraphics [width=\columnwidth] {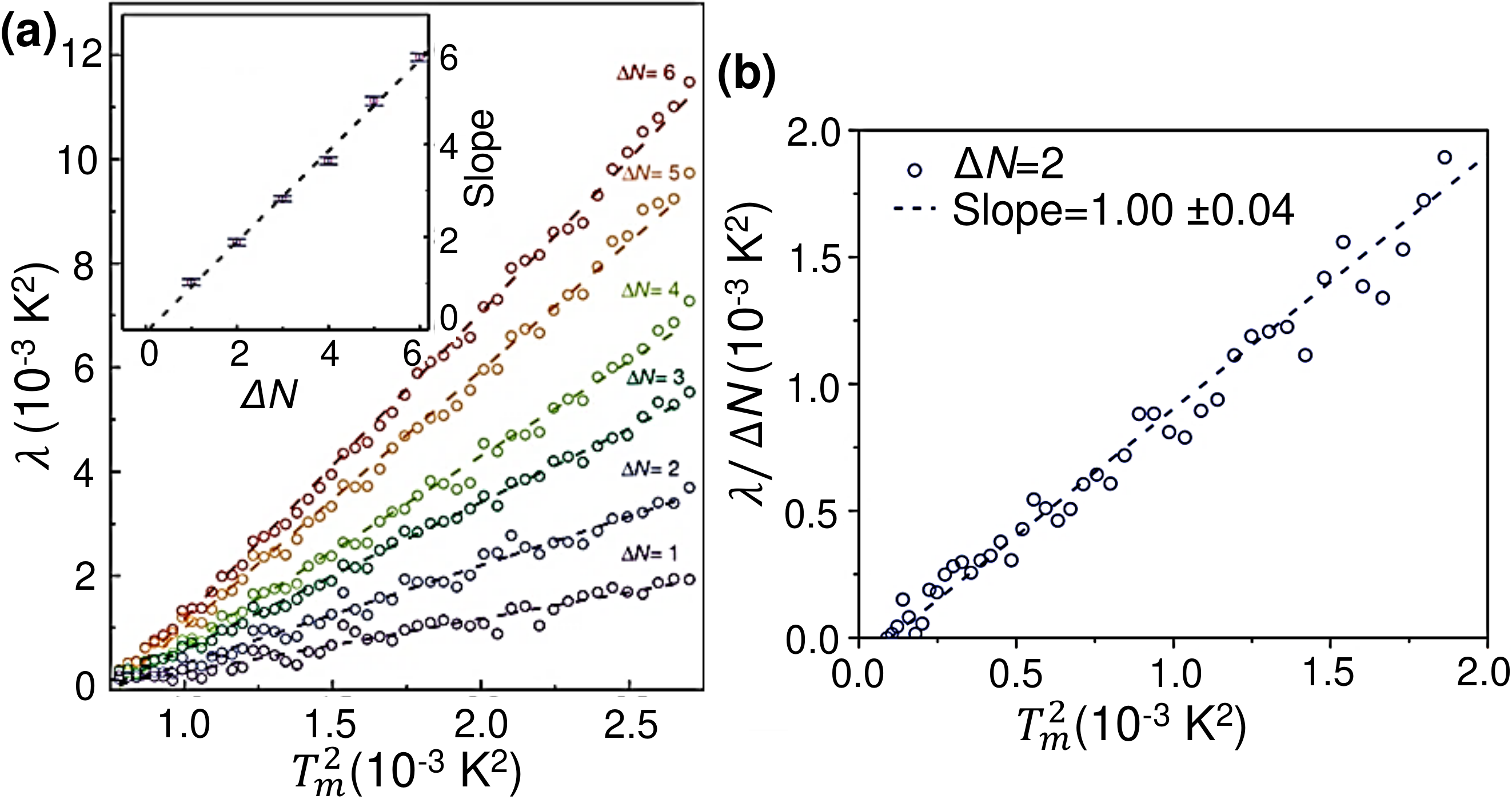}
	\caption{Measurements in the integer (a) and fractional (b) quantum Hall regimes with filling factors $\nu=2$ and $\nu=1/3$, respectively. (a) The normalized coefficient of the dissipated power $\lambda=\delta P/(G_{\rm Q}/2T)$ as a function of $T_{\rm m}^2$ for different configurations of $\Delta N=N_i-N_j$, where $N$ is the number of channels. The difference is presented in order to eliminate the $N$-independent contribution of the phononic heat current. Here, $\delta P$ is the difference between dissipated power at different $N$, $\delta P=\Delta P(N_i, T_{\rm m})-\Delta P(N_j, T_{\rm m})$, and $T_{\rm m}$ is the calculated temperature of the floating contact. The circles are showing the measured data and the dashed lines are linear fits to them. The slope of each set is shown in the inset as a function of $\Delta N$. The linear dependence has approximately unit slope ($0.98\pm 0.03$) confirming the quantum of thermal conductance for this integer state $\nu=2$. (b) The case of fractional state $\nu=1/3$. Same as (a) but here the difference of $\lambda$ between $N=4$ and $N=2$ normalized by $\Delta N$ as a function of $T_{\rm m}^2$. The slope of the linear fit (dashed line) to the measured data (circles) is very close to unity. Adapted from~\cite{Banerjee2017}.
		\label{Banerjee2017}}
\end{figure}

In the measurement performed by~\cite{Banerjee2017}, the value of the quantum of thermal conductance for different Hall states including integer and fractional states was verified. First, they confirmed the observations of~\cite{Jezouin601} in a similar setup in the integer states with filling factors $\nu=1$ and 2. Figure~\ref{Banerjee2017}(a) demonstrates the validity of quantized heat conductance at $\Delta NG_{\rm Q}$ for $\Delta N=1,~2,...,6$ channels with about 3\% accuracy (inset). The main result of the work is the observation of thermal conductance of strongly interacting fractional states. Figure~\ref{Banerjee2017}(b) shows that the thermal conductance is again a multiple of $G_{\rm Q}$ even for the (particle-like) $\nu=1/3$ fractional state, although the electrical conductance is normalized by the effective charge $e^*=e/3$. As a whole, the work covers both particle- and hole-like fractional states, testing the predictions of~\cite{Kane1997}.

As a final point for this Section we want to mention that there is a large number of further beautiful experiments on various heat transport effects performed in the quantum Hall regime. Not going into details of them here, we list a few of them for further reading:~\cite{Sueur2010,Altimiras2010,Banerjee2018,Sivre2018,Granger2009,Nam2013,Halbertal2016,Halbertal2017}.

\section{Photons}\label{Photons}
In this section we discuss transport by thermal microwave photons, presenting another important bosonic system to study in this context. 

\subsection{A ballistic photon channel}
\begin{figure}
	\centering
	\includegraphics [width=\columnwidth] {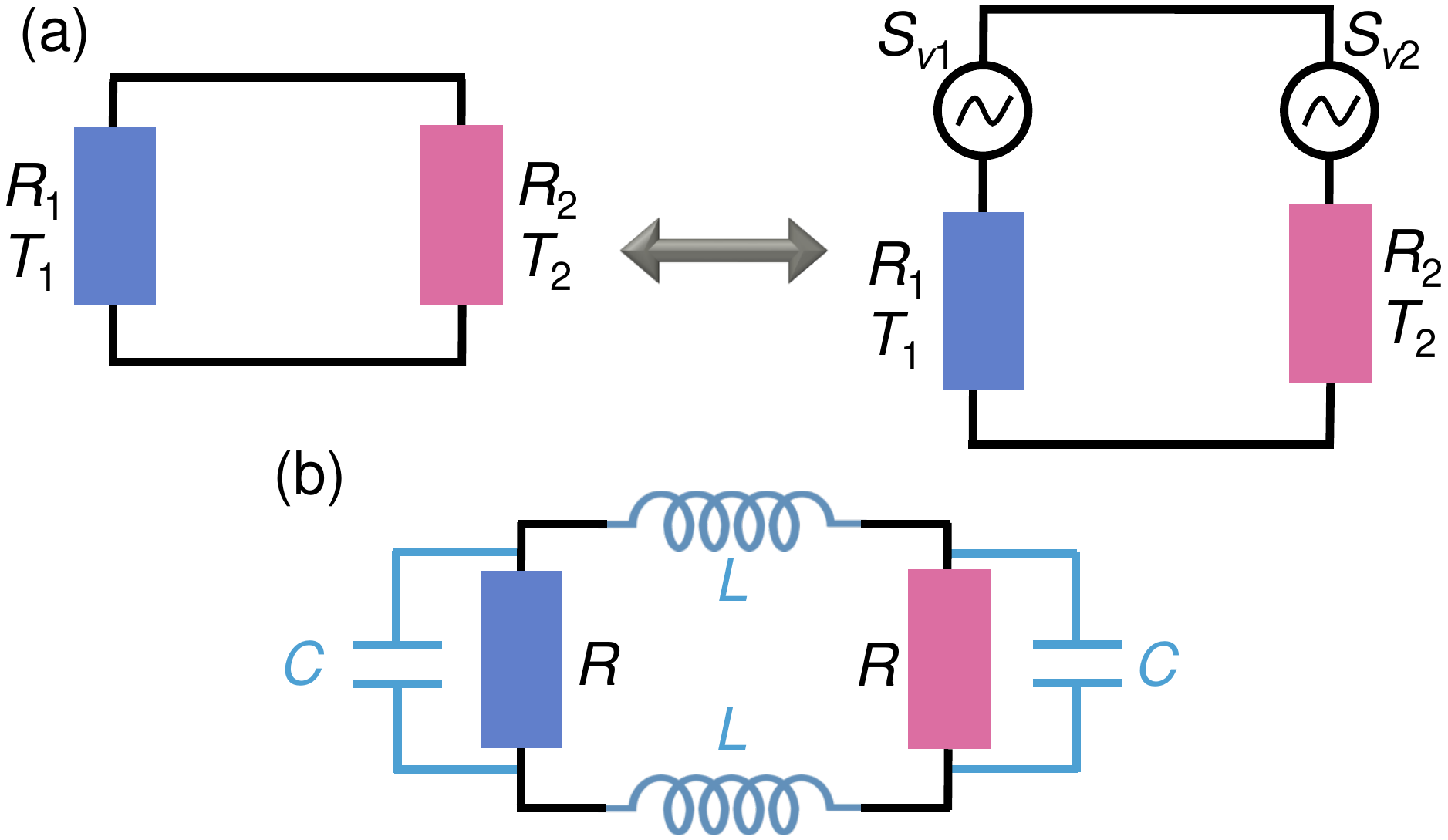}
	\caption{The setup of two resistors $R_1,R_2$ at temperatures $T_1, T_2$, respectively, interacting with each other via the respective thermal noises. We present the quantum version of the classical Johnson-Nyquist problem in the text with the associated radiative heat current. (a) The plain two-resistor heat exchange can be modeled by circuit approach where each resistor is accompanied with a thermal voltage noise source. The two sources are uncorrelated. (b) A realistic circuit includes inevitably reactive elements as well, as discussed in the text. These are added in the figure to allow for the analysis of the crossover between quantum and classical regimes upon varying the operating temperature and the physical system size.  
		\label{JohnsonNyquist}}
\end{figure}

The concept of microwave photon heat transport becomes concrete when describing it on a circuit level~\cite{Schmidt2004}. We start from a setup familiar from the century-old discussion by Johnson and Nyquist~\cite{Johnson1928,Nyquist1928}. There two resistors $R_1,R_2$ are directly coupled to each other as shown in Fig.~\ref{JohnsonNyquist}(a). They are generally at different temperatures $T_1$ and $T_2$. Each resistor then produces thermal noise with the spectrum $S_v(\omega)$ of Eq.~\eqref{positive2}, i.e., they can be considered as photon sources. First we consider that $R_1$ generates noise current $i_1$ on resistor $R_2$ as $i_1={v_1}/{(R_1+R_2)}$. Then the spectral density of current noise is $S_{i_1}(\omega)=(R_1+R_2)^{-2}S_{v_1}(\omega)$. The voltage noise produced by resistor $R_i$, $i=1,2$ is $S_{v_i}(\omega)=2R_i\hbar\omega/(1-e^{-\beta_i\hbar\omega})$, for $i=1,~2$. The power density produced by noise of $R_1$ and dissipated in resistor $R_2$ is then
$S_{P_2}(\omega)=\frac{R_2}{(R_1+R_2)^2}S_{v_1}(\omega)$.
The corresponding total power dissipated in resistor $R_2$ due to the noise of resistor $R_1$ is 
\begin{eqnarray} \label{e6}
	P_{2}&&=\int_{-\infty}^{\infty}\frac{d\omega}{2\pi}S_{P_2}(\omega)\nonumber\\&&=\frac{4R_1R_2}{(R_1+R_2)^2}\int_{0}^{\infty}\frac{d\omega}{2\pi}\hbar\omega[n_1(\omega)+\frac{1}{2}],
\end{eqnarray}
The net heat flux from 1 to 2, $P_{\rm net}$, is the difference between $P_{2}$ and $P_{1}$, where $P_{1}$ is the corresponding power produced by $R_2$ on $R_1$ by the uncorrelated voltage (current) noise described similarly. Thus
\begin{equation} \label{power1}
	P_{\rm net}=\frac{4R_1R_2}{(R_1+R_2)^2}\frac{\pi k_{\rm B}^2}{12\hbar}(T_1^2-T_2^2).
\end{equation}
Note that the integrals for $P_{1}$ and $P_{2}$ separately (see Eq.~\eqref{e6}) would lead to a divergence due to the zero point fluctuation term, but since these fluctuations cannot transport energy, this term cancels out in the physical net power Eq.~\eqref{power1}. We find that for a small temperature difference with $T_1=T_2\equiv T$
\begin{eqnarray} \label{G}
	G_\nu=\frac{dP_{\rm net}}{dT_1}|_{T}=\frac{4R_1R_2}{(R_1+R_2)^2}\frac{\pi k_{\rm B}^2}{6\hbar}T,
\end{eqnarray}
which is equal to the quantum of heat conductance 
\begin{equation} \label{gq}
	G_\nu=G_Q
\end{equation}
for $R_1=R_2$. For a general combination of resistance values the factor
\begin{equation}\label{ratio}
	r=\frac{4R_1R_2}{(R_1+R_2)^2}
\end{equation}
represents a transmission coefficient. The circuit model for heat transport can be generalized to essentially any linear circuit composed of reactive elements and resistors as has been done, e.g., in~\cite{Pascal2011,Thomas2019}.

\subsection{Circuit limitations of the ballistic picture}\label{Photons validity}
What are the physical conditions for the experiment in a circuit to yield thermal conductance that is governed by $G_{\rm Q}$? Certainly Johnson-Nyquist work~\cite{Johnson1928,Nyquist1928} was out of this domain, as well as an elegant more recent experiment by Ciliberto~\cite{Ciliberto2013}. The necessary key ingredients for "quantumness" are that the experiment combines low temperatures and physically small structures. More quantitatively, the realistic circuit is never presented fully by the simple combination of two resistors, but, insead the full picture of it includes also inevitable reactive elements. A way of describing a more realistic circuit~\cite{Golubev2015} is to include a parallel capacitance and series inductance in the basic circuit, as shown in Fig.~\ref{JohnsonNyquist}(b). The point is that electromagnetics tells us that an order of magnitude estimate for capacitance is given by $C\sim \epsilon \ell$ and inductance by $L \sim \mu_0 \ell$, where $\ell$ is the overall linear dimension of the circuit and $\epsilon$ and $\mu_0$ are the permittivity and permeability of the medium. In order to observe the pure quantum thermal conductance, one needs to have a ballistic channel between the resistor baths, which in the present case means that the series inductor presents a small impedance and the parallel capacitance a large impedance. These both are to be compared to the resistances in the circuit, at all relevant frequencies meaning up to $\omega_{\rm th}=k_{\rm B}T/\hbar$, the thermal cut-off of the resistor at temperature $T$. In form of simple inequalities we then need to require $\omega_{\rm th}L\ll R\ll (\omega_{\rm th}C)^{-1}$, and based on our arguments above this transforms into
\begin{equation}\label{ineq}
	\epsilon\ell k_{\rm B} T R/\hbar \ll 1, \,\, {\rm and}\,\,\mu_0 \ell k_{\rm B}T/(\hbar R) \ll 1. 
\end{equation}

It is now easy to verify the statements in the beginning of this subsection. We assume for simplicity a typical value for a resistance used in some experiments, $R=100$ $\Omega$. If we take a mesoscopic circuit with $\ell = 100$ $\mu$m at a low temperature, $T=100$ mK, we find that $\epsilon\ell k_{\rm B} T R/\hbar \approx \mu_0 \ell k_{\rm B}T/(\hbar R) \approx 0.01$ satisfying the conditions in Eqs. \eqref{ineq}. On the other hand a $\ell = 0.1$ m macroscopic circuit at room temperature $T=300$ K yields $\epsilon\ell k_{\rm B} T R/\hbar \approx \mu_0 \ell k_{\rm B}T/(\hbar R) \approx 3\times 10^4$, far in the classical regime. Some of those conditions can be avoided in a low temperature transmission line circuit~\cite{M.Partanen2016} as will be discussed below.

\subsection{Experiments on heat mediated by microwave photons}\label{Photons Experiments}
We modelled in Section~\ref{Photons} the heat emitted by a resistor and absorbed by another one in an otherwise dissipationless circuit. It was shown~\cite{Schmidt2004} that this heat carried by microwave photons behaves as if the two resistors were coupled by a contact whose ballisticity is controlled by the impedances in the circuit. Ideally two physically small and identical resistors at low temperatures can come very close to the ballistic limit with thermal conductance approaching $G_Q$. Motivated by this observation, several experiments assessing this result were set up during the past two decades~\cite{Meschke2006,Timofeev2009,M.Partanen2016}. They were all performed essentially in the same scenario: the resistors are normal metallic thin film strips with sufficiently small size such that their temperature varies significantly in response to typical changes of power affecting them. The electrical connection between the resistors is provided by superconducting (aluminum) leads, whose electronic heat conductance is vanishingly small at the temperature of operation, see Section~\ref{Experimental Thermal}. In one of the experiments (\cite{Meschke2006}) the superconducting lines are interrupted by a SQUID that acts as a tunable inductor providing a magnetic flux controlled valve of photon mediated heat current. All these experiments are performed at $T \sim 0.1$ K, far below $T_C \approx 1.4$ K of aluminum. Temperatures are controlled and monitored by biased NIS tunnel junctions. 

\begin{figure}
	\centering
	\includegraphics [width=\columnwidth] {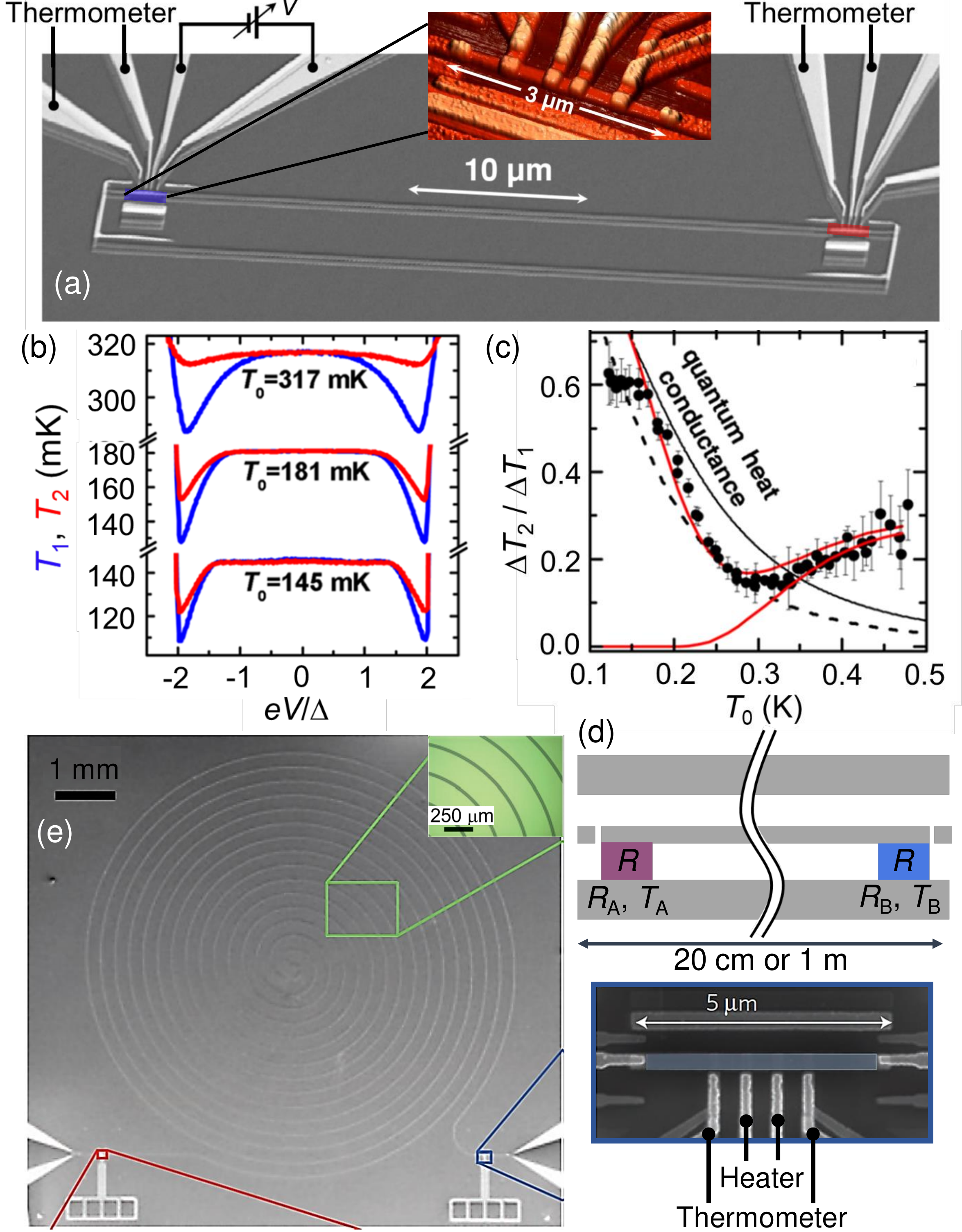}
	\caption{Quantum-limited heat conduction over microscopic and macroscopic distances. (a) The scanning electron micrograph (SEM) of two AuPd resistors at a distance of $50~\mu$m are connected via Al superconducting lines into a loop to match the impedance between them to reach the full quantum of heat conductance. NIS probe junctions are used to apply Joule heat and to measure the island temperature. (b) Measured local island temperature, $T_1$ (blue) and remote one $T_2$ (red) as functions of applied bias voltage on cooler/heater NIS junction pairs at different bath temperatures $T_0$. The drop for $T_1$ is naturally stronger than that for $T_2$. (c) The measured relative temperature drops (symbols) $\Delta T_2/\Delta T_1$ against $T_0$ at the optimum cooling bias voltage obtained from data like those in (b). The descending solid and dashed black lines are obtained from the linearized thermal model, which is simply given by $\Delta T_2/\Delta T_1=G_x/(G_x+G_{\rm th,2})$ (see Fig.~\ref{epset}(b)) at low temperatures. The thermal model is that shown in Fig.~\ref{epset}(b). The electron-phonon constant considered to be $\Sigma_{\rm AuPd}=2\times 10^9~{\rm WK}^{-5}{\rm m}^{-3}$ and $\Sigma_{\rm AuPd}=4\times 10^9~{\rm WK}^{-5}{\rm m}^{-3}$ for the solid and dashed lines, respectively. The remaining red solid lines are the results of the numerical thermal model. (d) A superconducting transmission line terminated at the two ends by normal-metal resistances $R_{\rm A}$ and $R_{\rm B}$ at two different electron temperatures $T_{\rm A}$ and $T_{\rm B}$, respectively, is schematically shown. (e) SEM image of an actual device, where the length of the coplanar waveguide (transmission line), made out of Al, is either 20~cm or 1~m and has a double-spiral structure. The SEM image of the zoom-out of one of the resistors (made out of either AuPd or Cu) with a simplified measurement scheme is shown in the bottom right of the figure. (a-c) Adapted from~\cite{Timofeev2009} and (d-e) adapted from~\cite{M.Partanen2016}.
		\label{test-timofeev}}
\end{figure}

The experiment of Timofeev et al.~\cite{Timofeev2009} was designed to mimic as closely as possible the basic configuration of Fig.~\ref{JohnsonNyquist}(a) with a superconducting (Al) loop. In this case the distance between the resistors was about 50 $\mu$m and the temperatures of both the heated (or cooled) source and the drain resistor were measured. The experiment (Fig.\ref{test-timofeev}(a-c)) demostrates thermal transport via the electronic channel, i.e. quasiparticle thermal transport~\cite{Bardeen1959} described in Section~\ref{Experimental Thermal}, at temperatures exceeding $\sim 250$~mK. The result in this regime is well in line with the basic theory, taken the dimensions and material parameters of the aluminum leads. More importantly, below about 200~mK the photon contribution kicks in. In the loop geometry it turns out that the temperatures of the two resistors follow each other closely at lowest bath temperatures, yielding thermal conductance given by $G_{\rm Q}$. Some uncertainty remains about the absolute value of $G_\nu$ since the precise magnitude of the competing electron-phonon heat transport coefficient $\Sigma$ remained somewhat uncertain. The measurement was backed by a reference experiment, where a similar sample as that described above was measured under the same conditions and fabricated in the same way. This reference sample lacked intentionally one arm of the loop leading to poor matching of the circuit in the spirit discussed in Section~\ref{Photons validity}. In this case the quasiparticle heat transport prevails as in the matched sample, but the photon $G_\nu$ is vanishingly small, confirming, one could say even quantitatively, the ideas presented about the heat transfer via a non-vanishing (reactive) impedance. 

The experiment described above was performed on a structure with physical dimensions not exceeding 100 $\mu$m. A natural question arises: is it possible to transport heat over macroscopic distances by microwave photons, like radiating the heat away from the whole chip? This could be important for example in quantum information applications (for superconducting qubit realizations, see~\cite{Kjaergaard2020}). This question was addressed experimentally by~\cite{M.Partanen2016} (Fig.\ref{test-timofeev}(d-e)), who placed the two resistors at a distance of about 10~mm, i.e. about 100 times further away from each other as compared to what was done earlier. Furthermore, the connecting line between the two baths was a 1~m long meander made of a superconductor, acting as a transmission line. Such a coplanar line has typically an impedance of about 50~$\Omega$ irrespective of its length, thus potentially supporting the heat transport even over large distances. The thermal conductance was measured as in Timofeev et al.~\cite{Timofeev2009}, with similar results proving the hypothesis of photon transport over macroscopic distances. These experiments may open the way for practical heat transport schemes in microwave circuits.
\begin{figure}
	\centering
	\includegraphics [width=\columnwidth] {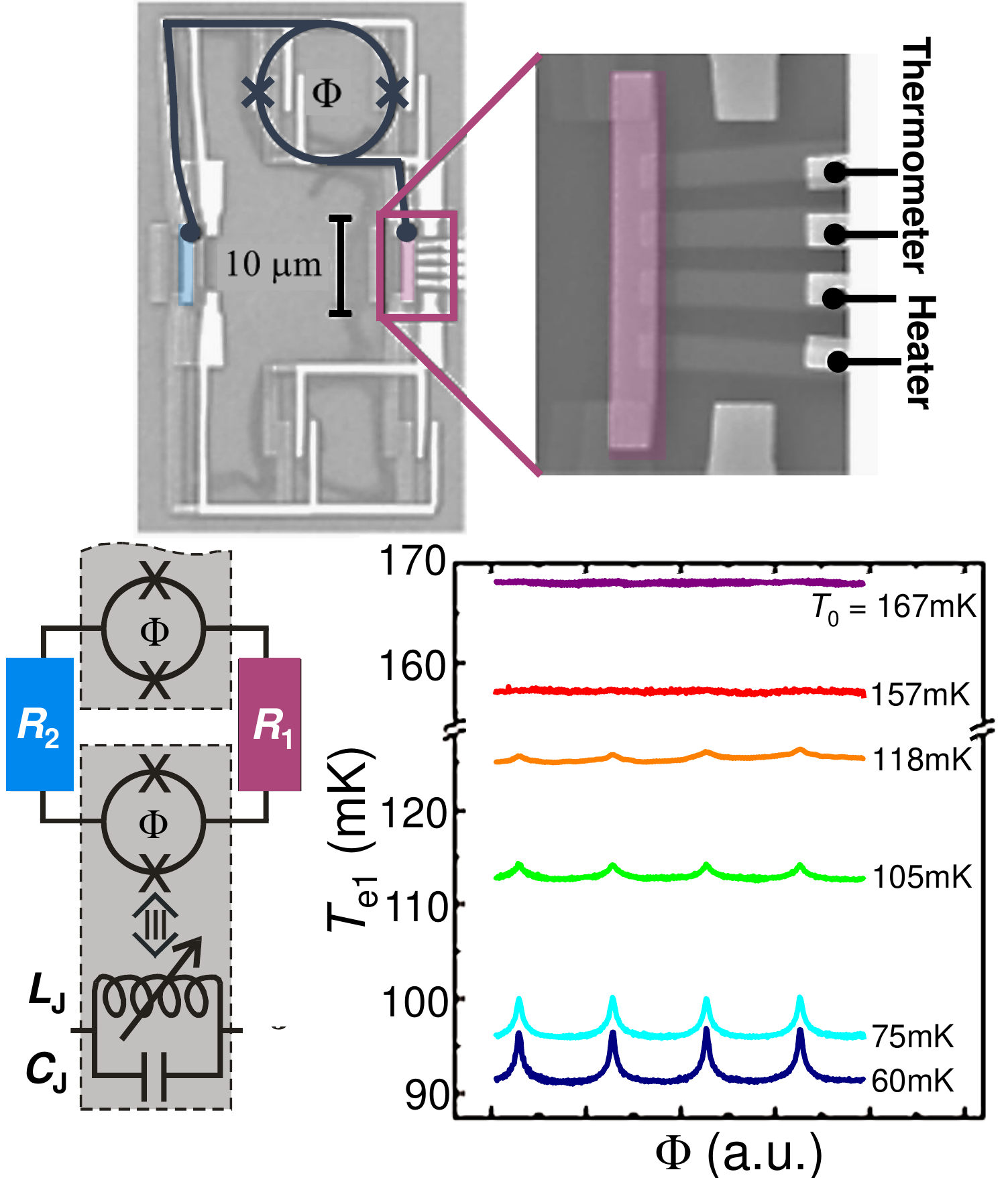}
	\caption{Photon thermal transport controlled by magnetic flux. (top left) The SEM of the device, where two nominally identical normal-metal resistors $R_1$ and $R_2$ (colored in blue and red on the left and right, respectively), made out of AuPd, are connected to each other via two aluminum superconducting leads, interrupted by a DC-SQUID (a superconducting loop with two Josephson junctions shown by the cross sign) in each line. The SQUIDs again serve as thermal switches between the resistors and can be controlled by external magnetic flux $\Phi$. (top right) Magnified view of $R_1$,  connected to four NIS tunnel junctions to the right and to two NS contacts at top and bottom to allow thermometry and Joule heating. (bottom left) Schematic illustration of the electrical model of the actual sample. (bottom right) Measured flux modulation of the electron temperature $T_{e1}$ in $R_1$ as a function of applied magnetic flux $\Phi$ for different values of bath temperature $T_0$. The modulation decreases monotonically by increasing $T_0$ because of stronger coupling to the phonon bath. The maxima in $T_{e1}$ correspond to the weakest electron-photon coupling at half-integer values of $\Phi/\Phi_0$ where $\Phi_0=h/2e$ is the magnetic flux quantum. Adapted from~\cite{Meschke2006}.
		\label{Meschke-Nat}}
\end{figure}

\section{Tunable quantum heat transport}\label{Tunable quantum}
In this section we describe quantum systems where heat transport is controlled by either magnetic or electric field to achieve useful functional operation. These devices include heat valves, heat interferometers, thermal rectifiers and circuit refrigerators. Mesoscopic structures provide an option to control currents by external fields. Concerning charge currents, SQUIDs~\cite{Tinkham2004} and single-electron transistors~\cite{Averin1991} provide hallmark devices in this context, where magnetic field (flux in a superconducting loop) and electric field (gate voltage), respectively, are the parameters that control the current. 

The first experiment on heat transport by thermal microwave photons~\cite{Meschke2006} was realized in a setup where a SQUID is used as a heat valve. The experiment depicted in Fig.~\ref{Meschke-Nat} shows two metallic (AuPd) resistors at the distance of few tens of $\mu$m from each other, connected by superconducting (Al) lines. The loop is interrupted in each arm by a SQUID, whose flux can be controlled by the common external field for both of them. The thermal model of Fig.~\ref{epset}(b) applies to this circuit. In the experiment only the heated resistor's temperature $T_{e1}$ was measured. The panel at bottom right displays the magnetic flux dependent variation of temperature $T_{e1}$ at different bath temperatures $T_0$ under a constant level of heating. At bath temperatures well above 100~mK, the flux dependence vanishes, since the inter-resistance thermal conductance by photons, $\propto T$, is much weaker than the conductance to the phonon bath, $\propto T^4$. On the contrary, towards low temperatures below 100~mK, the electron temperature $T_{e1}$ varies with magnetic flux as the inter-resistor coupling becomes comparable to the bath-coupling, demonstrating the photonic thermal conductance. Moreover, the magnitude of the thermal conductance was shown to follow from the circuit model presented in Section~\ref{Thermal Principles} quantitatively, when applied to the current setup. Among other things, the data and this calculation predicted that at $T_0=60$~mK the maximum value of thermal conductance with zero flux in the SQUID (i.e. with minimum Josephson inductance) was $\sim 50$\% of $G_{\rm Q}$.

\begin{figure}
	\centering
	\includegraphics [width=\columnwidth] {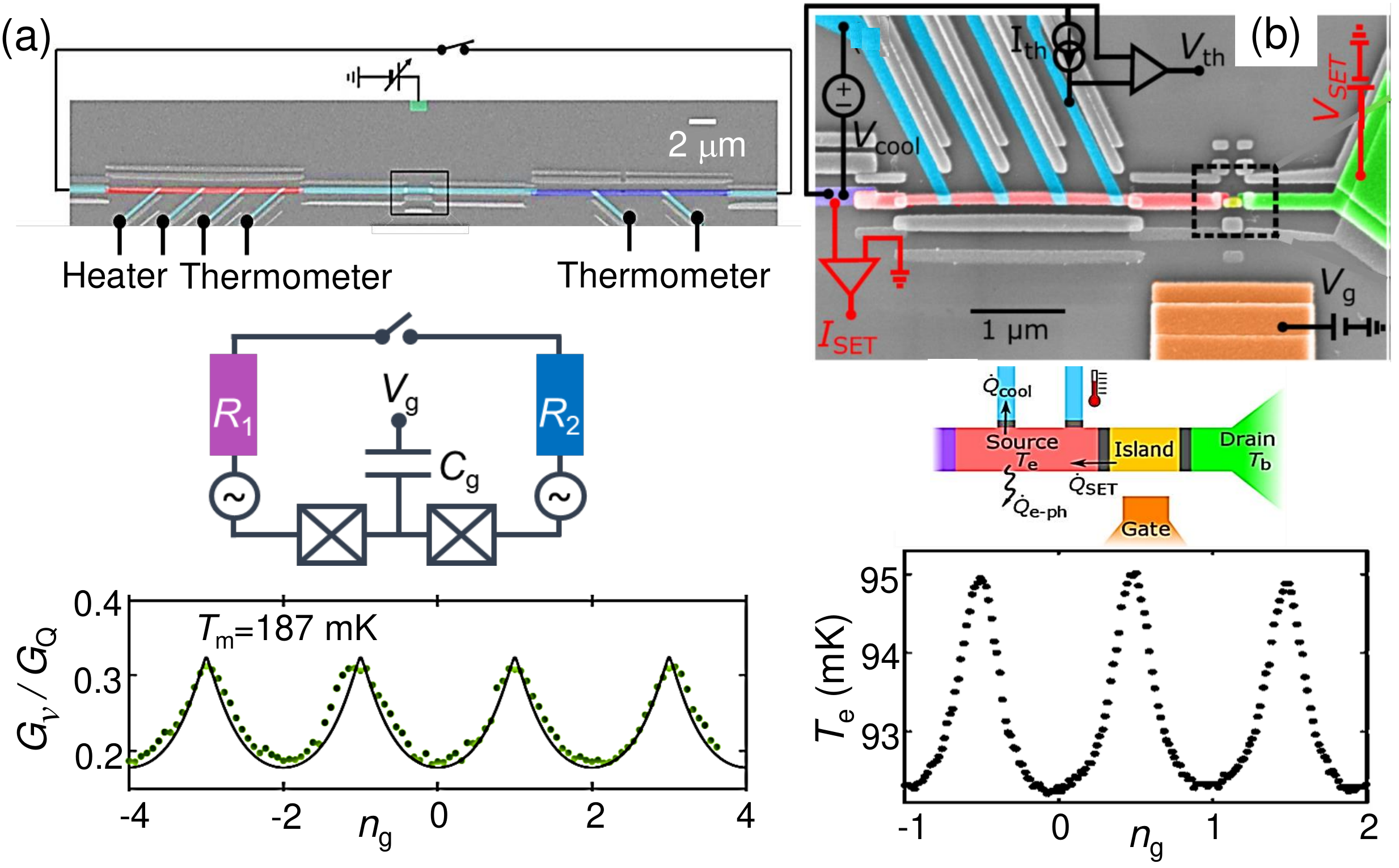}
	\caption{(a) The SEM (top) and the equivalent electrical circuit diagram (middle) of the device including a Cooper-pair transistor coupled to two normal-metal resistors $R_1$ and $R_2$ at temperatures $T_1$ and $T_2$, respectively. (bottom) Measured thermal conductance (symbols) normalized by $G_{\rm Q}$ as a function of gate charge $n_{\rm g}=C_{\rm g}V_{\rm g}/e$ at $T_{\rm m}=(T_1+T_2)/2$. The solid line indicates the theoretical expectation. (b) The SEM image (top) and the schematic realization (middle) of the device consisting of a single-electron transistor and the heat transport measurement setup. The black circuit on the top left corner displays the heat transport setup. (bottom) Measured gate dependence of the electronic temperature $T_{\rm e}$ of the source island when it is lower than the bath temperature $T_0$. (a) Adapted from~\cite{Maillet2020} and (b) adapted from~\cite{Dutta2017}.
		\label{Bivas-Olivier}}
\end{figure}

Photonic heat current was controlled by magnetic field in the previous example. A dual method is to apply electric field as a control as indicated in Fig.~\ref{Bivas-Olivier}(a). This procedure was realized in the experiment recently~\cite{Maillet2020}, where the superconducting loop is interrupted by a Cooper-pair transistor ("charge qubit"~\cite{Nakamura1999}). In this setup, the Josephson coupling is tuned by the gate voltage. The thermal model of the experiment is pretty much as before, only the Josephson element with its control field is different. Electric field by gate voltage is in general easier to apply, especially locally on the chip, as compared to local magnetic flux. As shown by Fig.~\ref{Bivas-Olivier}(a), the device demonstrated gate-dependent modulation of heat current. Its overall magnitude is consistent with the modeling of the circuit: at maximum thermal conductance $G_{\nu}\approx 0.35 G_{\rm Q}$ was achieved.

As to the heat currents in single-electron circuits, similar control principles apply, in general. An early experiment to control heat flow by a gate voltage in a single-electron transistor formed of NIS junctions was performed by~\cite{Saira2007}. The results on temperature of the system were confirmed quantitatively by a model employing standard single-electron tunneling theory and heat balance equation on the measured central island of the transistor. Heat transport via a fully normal metallic single-electron transistor was measured by~\cite{Dutta2017}. Results of this experiment are shown in Fig.~\ref{Bivas-Olivier}(b). The heat current between the source and drain with a temperature bias applied across was carried by electrons and modulated by the gate voltage such that the observed thermal conductance and simultaneously measured electrical conductance go hand in hand. Yet, deviations from the Wiedemann-Franz law due to Coulomb blockade and quantum tunneling were observed in agreement with the theory~\cite{Kubala2008}. Recently, deviations from Wiedemann-Franz law were observed in a similar setup with a semiconducting InAs quantum dot transistor~\cite{Majidi2021}.

\subsection{Electronic quantum heat interferometer}\label{Electrons Electronic}
\begin{figure}
	\centering
	\includegraphics [width=\columnwidth] {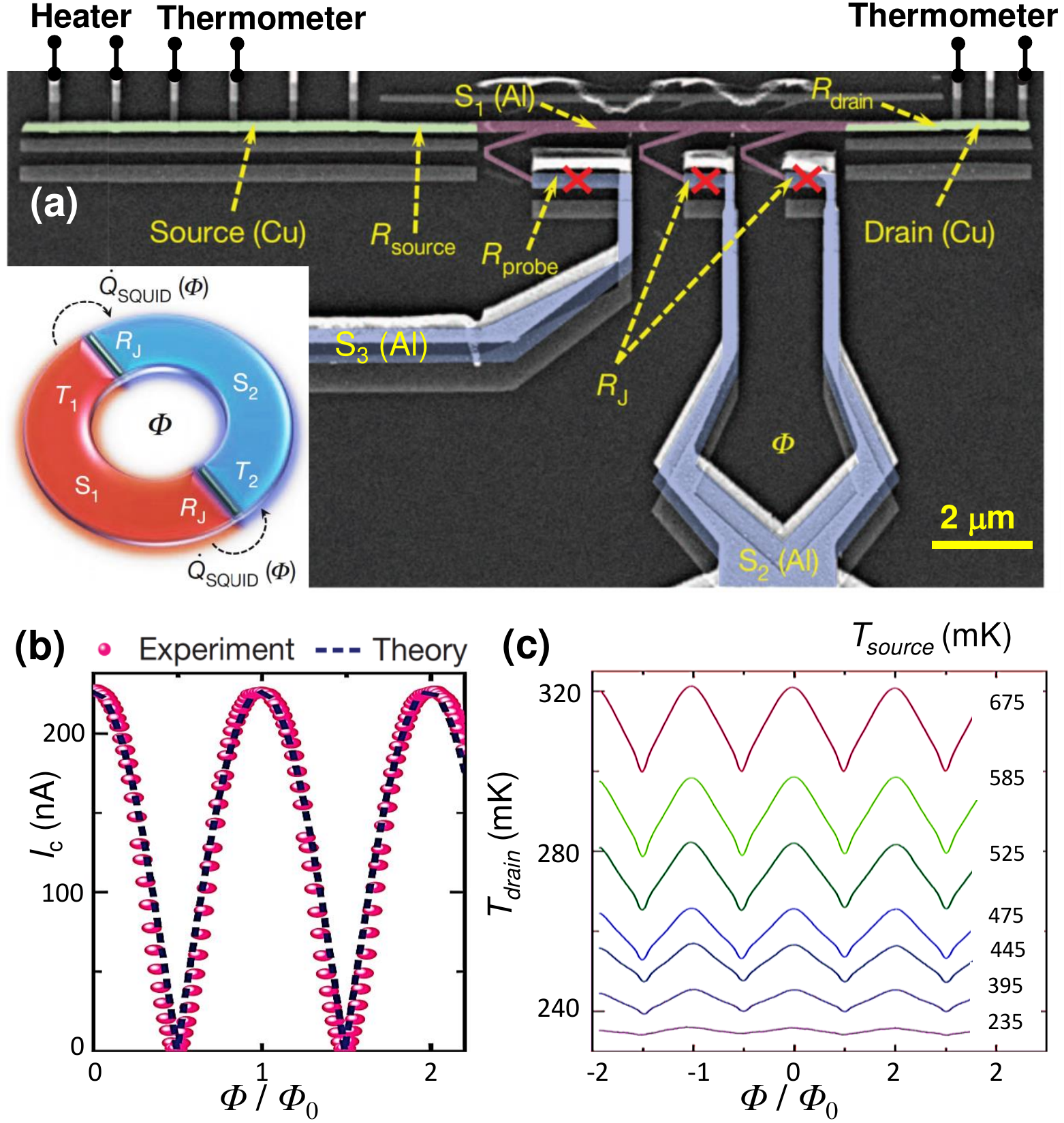}
	\caption{Josephson heat interferometer. (a) The SEM of the device. The source and drain electrodes made out of Cu are connected to an Al island (${\rm S_1}$) through two AlO$_{\rm x}$ tunnel barriers. ${\rm S_1}$ is connected sideways to a DC-SQUID which is terminated to a large-volume lead ${\rm S_2}$ (Al) for thermalization and to an Al tunnel probe (${\rm S_3}$) for independent SQUID characterization. NIS junctions in source and drain are used to heat and monitor the temperature of each island. Red crosses indicate the Josephson junctions. The core of the device (SQUID) is shown schematically in the inset. Two identical superconductors with different temperatures are connected with the two tunnel junctions of the SQUID. Applying magnetic flux $\Phi$ the heat current $\dot{Q}_{\rm SQUID}(\Phi)$ from hot to cold is varying. (b) Maximal charge current of the SQUID $I_{\rm c}$ as a function of $\Phi$ at 240 mK bath temperature. The dashed line presents the theoretical result assuming $0.3\%$ asymmetry in the junctions, and symbols are experimental data. (c) Flux modulation of $T_{\rm drain}$ related to heat current measured at different $T_{\rm source}$ values. Here the bath temperature is fixed at 235~mK. Adapted from~\cite{Giazotto2012}.
		\label{Giazotto-interfero}}
\end{figure}

Another quantum interference experiment on heat current by electrons was performed by~\cite{Giazotto2012}, shown in Fig.~\ref{Giazotto-interfero}. They used a magnetic field controlled SQUID as an interferometer. They could independently measure the electrical and heat transport via the device. For the latter, the SQUID was placed between two mesoscopic heat baths and the heat current was measured with the principle that was depicted in Fig.~\ref{epset}(b). The measurement was performed in a temperature regime exceeding that described in Section~\ref{Experimental Thermal} such that $T$ is high enough for the superconductor to have a substantial equilibrium quasiparticle population (i.e. not all electrons are paired). In this regime the superconductor as such can support heat current, and heat interference across the Josephson junctions of the SQUID becomes possible. This work addressed experimentally for the first time a half-a-century old proposal and theory~\cite{Maki1965} which has been followed by several works more recently~\cite{Guttman1998,Guttman1997,Zhao2003,Golubev2013}. It also demonstrated potential of electronic caloritronics in superconducting circuits. 

\begin{figure}
	\centering
	\includegraphics [width=\columnwidth] {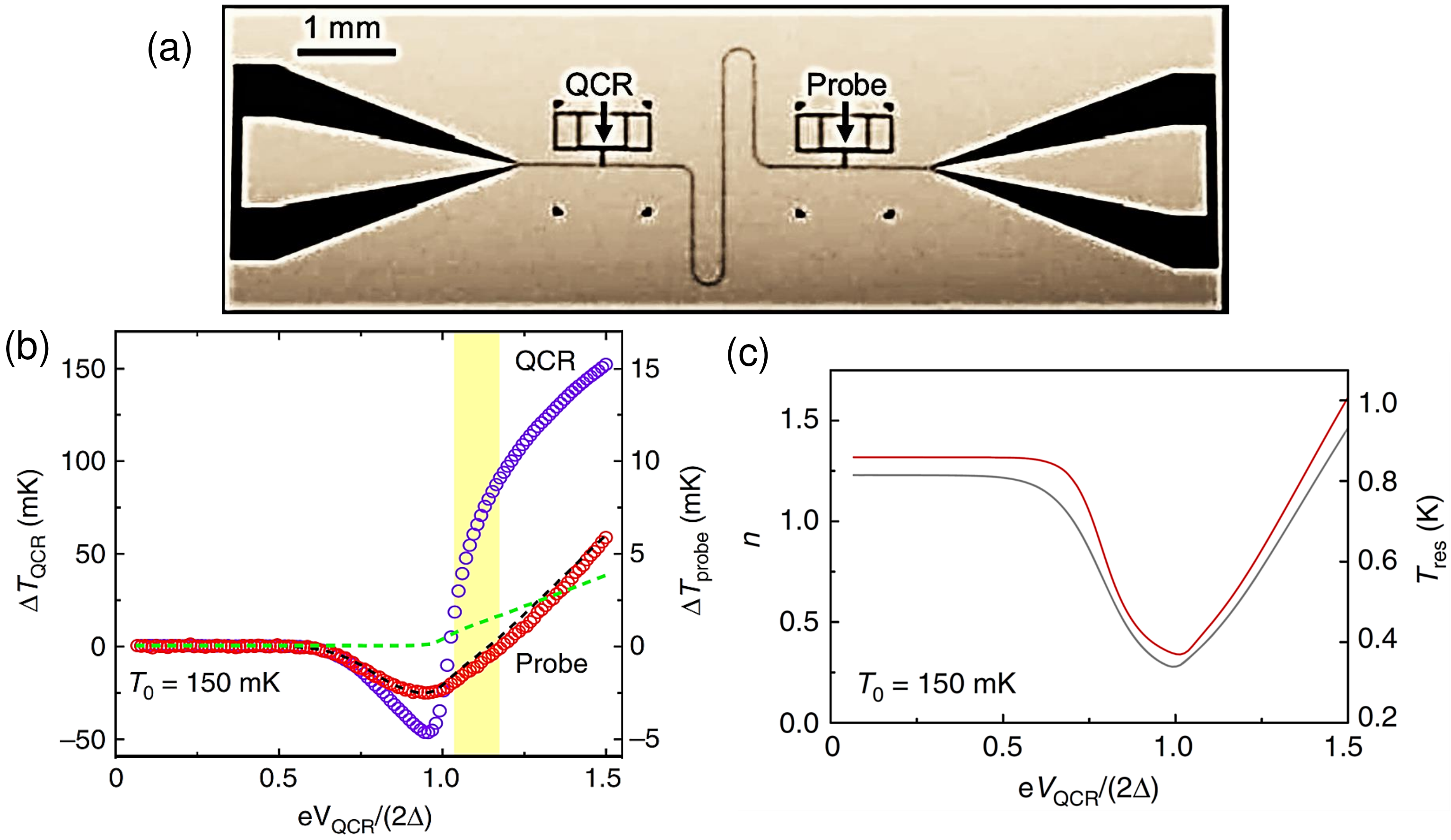}
	\caption{Quantum circuit refrigerator (QCR). (a) Optical micrograph of a sample, where a superconducting coplanar-waveguide resonator is in the center coupled to a QCR and a probe resistor indicated  by the arrows. (b) Measured changes in the electron temperature of QCR, $\Delta T_{\rm QCR}$ with purple circles, and the probe resistor, $\Delta T_{\rm probe}$, with red circles, as functions of the refrigerator operation voltage $V_{\rm QCR}$. The black dashed line, following the probe data closely, and green dashed line show the given theoretical results on $\Delta T_{\rm probe}$, including and excluding photon-assisted tunneling, respectively. (c) Average number of photons $n$ and the corresponding effective temperature of the resonator $T_{\rm res}$ shown with the two solid lines in red and grey, based on the thermal model introduced in~\cite{Tan2017}. Adapted from~\cite{Tan2017}.
		\label{QCR}}
\end{figure}

\begin{figure*}
	\centering
	\includegraphics [width=\textwidth] {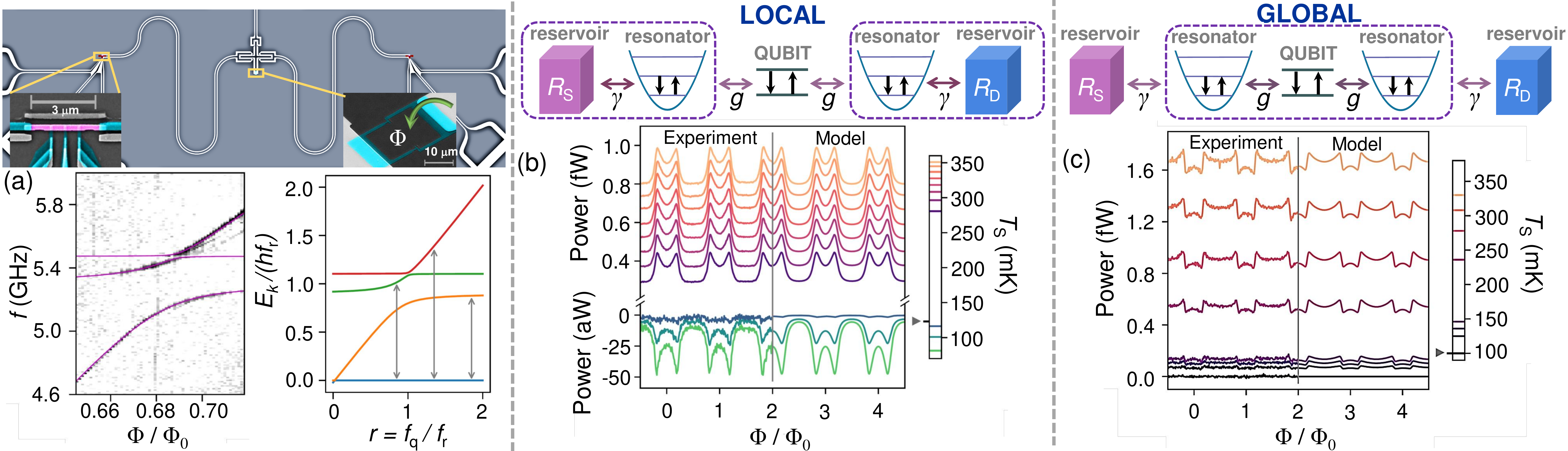}
	\caption{Quantum heat valve: the local and global regimes. (a) The top panel shows the quantum heat valve device where transmon-type superconducting qubit in the center is capacitively coupled to two similar $\lambda/4$ coplanar waveguide resonators made out of Nb. Each resonator is terminated by a normal-metal resistor (Cu) acting as a reservoir. The insets display the SEM images of the SQUID which can be controlled by magnetic flux $\Phi$, and one of the two reservoirs (pink, in the center) whose temperature is monitored and controlled by NIS probes (Cu-AlO$_{\rm x}$-Al). The lower panels show the two-tone transmission spectroscopy (left) and the corresponding theoretical positions (right) for the structure shown on top without resistors. (b,c) The top panels show the schematic illustrations of the two regimes. The resulting heat transport data are shown in the lower panels, where the flux dependence of the heat current in the drain reservoir in the two regimes is measured. Each trace corresponds to a different temperature $T_{\rm S}$ of the source reservoir shown in the legend bar. Adapted from~\cite{Ronzani2018}.
		\label{QHV}}
\end{figure*}

\begin{figure}
	\centering
	\includegraphics [width=\columnwidth] {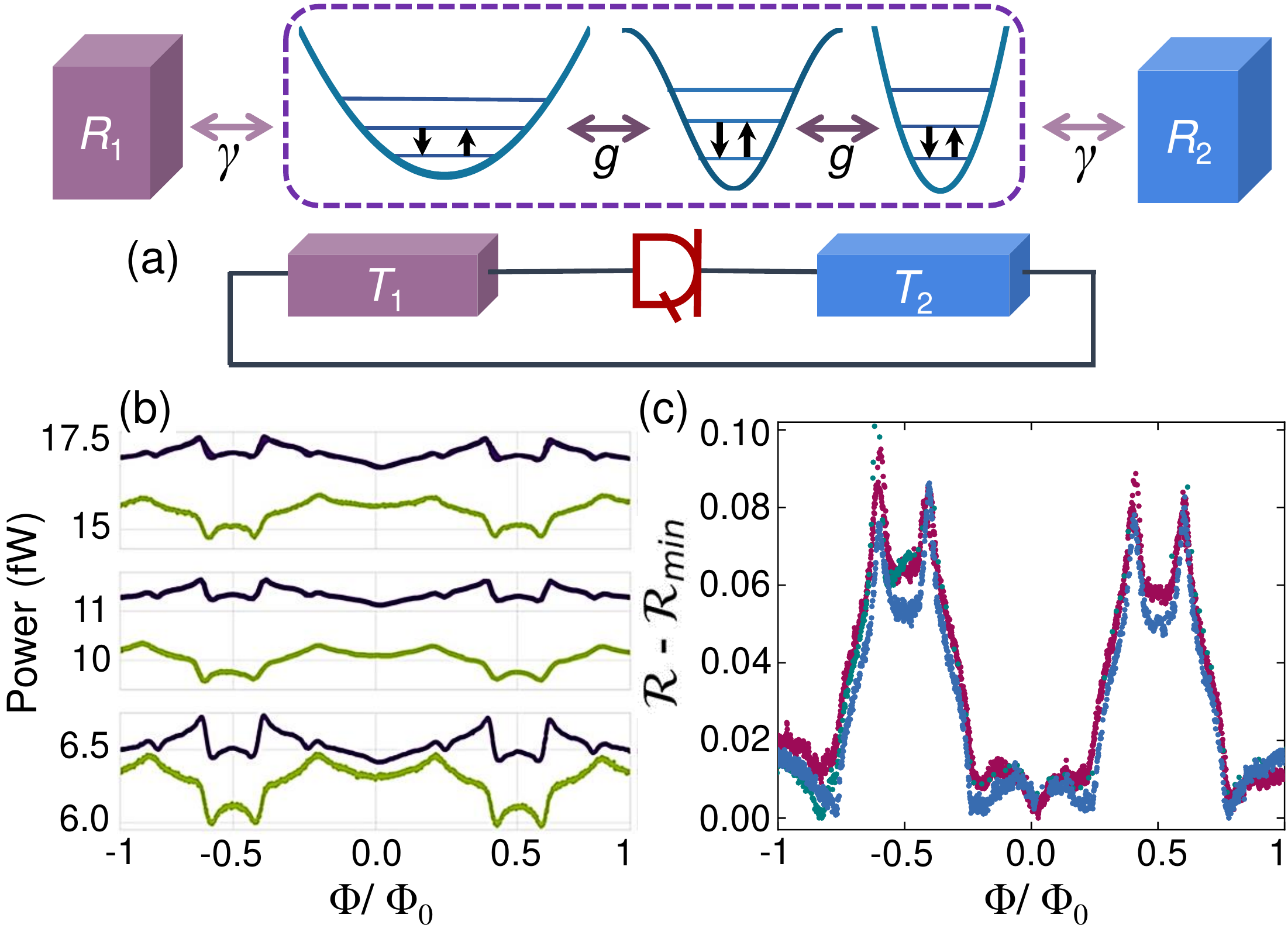}
	\caption{Heat rectification using a transmon qubit~\cite{Senior2020}. (a) A schematic illustration of a photon diode composed of an anharmonic oscillator (qubit), coupled to two $LC$ resonators with largely different resonance frequencies. The bottom panel in (a) shows a circuit view of the system, where we assign the quantum heat rectifier in the middle by the red symbol. (b) Dependence of source-drain heat current (power) as a function of magnetic flux in forward ($1\rightarrow 2$, dark line, $P^+$) and reverse ($2\rightarrow 1$, light line, $P^-$) directions under identical but opposite bias conditions at few different source temperatures (420, 400 and 380 mK from top to bottom). (c) Rectification ratio $\mathcal{R} = P^+/P^-$ (with its minimal value subtracted) as a function of magnetic flux for the data in (b) at the three temperatures of the source. Adapted from~\cite{Senior2020}.
		\label{Jorden-rec}}
\end{figure}

\subsection{Cooling a quantum circuit}
In the experiment performed by~\cite{Tan2017} photon-assisted tunneling serves the purpose of decreasing the number of microwave quanta in a superconducting quantum circuit, namely a coplanar wave resonator (harmonic oscillator). The optical micrograph of the sample presented in Fig.~\ref{QCR}(a) shows resistive elements inserted at the two ends of the resonator, acting as heat sinks for it. Figure~\ref{QCR}(b) displays the temperature of one of these resistors (Quantum Circuit Refrigerator, QCR), measured and controlled by NIS tunneling, effectively lowering and elevating the electronic temperature of it depending on the biasing of the cooler junction, see Section~\ref{Experimental Heat}. The other resistor (“Probe”) is passive but its temperature is likewise monitored. This temperature reacts weakly to the QCR temperature changes. The authors in~\cite{Tan2017} develop a thermal model based on which they extract the average number of photons in the resonator and the corresponding temperature $T_{\rm res}$. In this experiment $T_{\rm res}\approx 800$ mK far exceeds all other temperatures, most notably the electronic temperatures of the two resistors, $T_{\rm QCR}\approx T_{\rm probe}\approx 150$ mK even under no bias on the QCR. The model then predicts cooling of the resonator down to about 400~mK under optimal biasing conditions of the QCR~(Fig.~\ref{QCR}(c)). Based on the parameters given in the manuscript, one would estimate the resonator to have $T_{\rm res}\approx 200$ mK when the QCR is not biased. Indeed, in a later work~\cite{Masuda2018} resonator temperatures in the 200~mK regime were reported at zero bias. When biased, the NIS junctions operate as an incoherent microwave source. Then the mode temperature of the resonator can be driven even beyond 2.5~K, far above the temperature of the phonon and electron reservoirs of the system~\cite{Masuda2018}. This phenomenon was theoretically modelled by~\cite{Silveri2017} using photon-assisted tunneling of the biased, NIS-junctions as the environment. The effective temperature of the resonator is expected to be lifted to $\sim e V/(2 k_{\rm B})$ at bias voltage $V$.

\section{Quantum heat transport mediated by a superconducting qubit}\label{Photons Quantum}

In this section we introduce a superconducting qubit as an element that mediates heat by microwave photons between two baths. Different types of superconducting quantum bits, e.g. flux, charge and transmon qubits to mention a few common ones, are options in such devices~\cite{Clarke2008}. They feature different coupling options and strengths, as well as different degrees of anharmonicity in the Josephson potential to be discussed below. In the experiments of~\cite{Ronzani2018} transmon type qubits~\cite{Koch2007} were employed. This kind of a qubit has levels whose positions can be controlled by magnetic flux through the SQUID-loop. Transmon qubit is only weakly anharmonic, meaning that one typically needs to consider not only the two lowest levels (that form the actual qubit) but also the higher levels in this nearly harmonic potential. One should point out an important difference: although even weak anharmonicity is enough to address only the two lowest levels in a microwave-driven experiment, one needs, on the contrary, to consider higher levels as well, when the qubit sees a thermal bath with wide spectrum. Yet in a typical experiment to be described here, the separation of the levels is of the order of 0.5~K, meaning that the thermal population of the third level is already quite small at the low temperatures of the experiment, say, below 0.2~K ($\sim e^{-5} < 0.01$).

In this Section we present thermal transport experiments under the conditions where the qubit is not driven. Coherent properties of the qubit do not then play any important role. In the future the same devices will be driven by RF-fields and evolution of off-diagonal elements of the density matrix will develop as well. 

\subsection{Quantum heat valve} \label{Photons Heat}
Figure~\ref{QHV}(a) shows on the top a typical experimental configuration of heat control with qubit, taken from~\cite{Ronzani2018}. The energy separation of the transmon qubit (center) can be controlled by external magnetic flux $\Phi$. The qubit is coupled capacitively (coupling $g$) to two nominally identical superconducting coplanar wave resonators that act as $LC$ resonators with a resonant frequency of $\sim 5$~GHz each. For thermal transport experiments the $\lambda/4$ resonators are terminated by on-chip resistors which form the controlled dissipative elements in the circuit~\cite{Chang2019}. The dissipation is then given by the inverse of the quality factor of the resonator and can be quantified by another coupling parameter $\gamma$. In this circuit, which is called a quantum heat valve, heat is carried wirelessly (via capacitors) by thermal microwave photons over a distance of few millimeters from one bath to another. A schematic model of such a coupled circuit is shown in panels \ref{QHV}(b) and \ref{QHV}(c) on top. 

It turns out that the measurement of heat transport in such a circuit addresses some fundamental questions of open quantum systems~\cite{Levy_2014,Chiara2018,Hofer_2017,Rivas_2010,Hewgill2020,Magazz__2019,Aurell2019,Donvil2020,Donvil2018}. There are (at least) two possible ways of viewing the circuit, namely the local view (Fig.~\ref{QHV}(b)) and the global view (Fig.~\ref{QHV}(c)). In the local picture as we define it, the environment of the qubit is formed of the dissipative $LC$ resonator with Lorentzian noise spectrum centered around its resonance frequency. In this regime, which occurs when $\gamma \gg g$, the system acts indeed as a valve admitting heat current through only when the qubit frequency matches (within the range determined by the quality factor) the frequency of the resonators. This results in Lorentzian peaks in power centered at flux positions corresponding to the said matching condition, demonstrated by both experiment and theory, shown in the bottom panel of Fig.~\ref{QHV}(b). In the opposite limit, in the global regime, when $\gamma \ll g$, the situation is different. The combined system composed of the resonators and the qubit then make up a hybrid that interacts with bare environment formed of the two resistors. In this limit the hybrid quantum system has the energy spectrum shown in the bottom panel of Fig.~\ref{QHV}(a) exhibiting multilevel structure. This is shown by the basic calculated spectrum and the spectroscopic measurement on a structure similar to that in the top panel but in the absence of the resistive loads. The data in Fig.~\ref{QHV}(c) demonstrates results in the global regime, with experiment and theory developed in~\cite{Ronzani2018} in agreement with each other. This experiment is the first one to assess local-global crossover in the spirit of locating the Heisenberg cut between the quantum and classical worlds. In a recent theoretical analysis, the authors of this review analyzed the cross-over behavior between the two limiting regimes with the help of direct solution of the Schr\"odinger equation including an oscillator bath~\cite{Pekola2020}.

\subsection{Thermal rectifier}\label{Thermal rectifier}
In a symmetric structure as in Fig.~\ref{QHV}, there is naturally no directional dependence of heat transport between the two baths. However, heat current rectification becomes possible if one breaks the symmetry of the structure~\cite{Segal2005}. Heat recitification~\cite{Segal2005,Kargi2019,Ruokola2009,Motz2018,RieraCampeny2019,Goury2019,Sothmann2012,Sanchez2015,Bhandari2021,Iorio2021} can be quantified in different ways, but in general finite rectification means that the magnitudes of forward and reverse heat currents differ under identical but opposite temperature biasing conditions. There exist a few experiments on heat current rectification, e.g. on phonons in carbon nanotubes~\cite{Chang2006}, and electrons in quantum dots~\cite{Scheibner_2008}, mesoscopic tunnel junctions~\cite{MartinezPerez2015} and suspended graphene~\cite{Wang2017}. In~\cite{Senior2020} rectification was realized in a structure similar to that in Fig.~\ref{QHV} but by making the two resonators of unequal length: the two resonators had in this case frequencies 3~GHz and 7~GHz. An additional feature necessary for heat rectification is the non-linearity of the central element, which arises from the anharmonicity of the transmon Josephson potential. Figure~\ref{Jorden-rec} shows data from~\cite{Senior2020} where heat current through the structure is measured in forward and reverse directions under the same but opposite temperature biasing, respectively. Complicated flux dependence can be observed, but the main feature is that one reaches 10\% rectification at best and it depends strongly on the flux position determining the coupling asymmetry to the two baths. Quantitative analysis of the flux dependence is challenging and experiments in simpler setups would be welcome.  

\section{Heat current noise} \label{Heat current noise}
In this section we focus on fluctuations of currents, which are generally considered to be harmful, and something to get rid of. The synonym of fluctuations, noise, proposes this negative side of the concept of fluctuations. Noise typically determines the minimal detectable signal in a measurement, i.e. the resolution. Here we do not consider noise caused by the measurement apparatus or from other extrinsic sources, but we focus on intrinsic noise, due to fundamental quantum and thermal fluctuations. This noise, for instance in form of electrical current or heat current fluctuations determines the ultimate achievable measurement accuracy. But besides being a limiting factor of a measurement, noise can also serve as a signal to build on in order to realize a sensor: for instance, measurement of thermal current noise of a conductor provides one of the most popular and fundamental thermometers in use~\cite{Fleischmann2020}.

We already discussed current and voltage noise of a linear dissipative element in Section~\ref{Experimental Quantum}. Here we review the heat current noise, both classical and quantum, see e.g.~\cite{Crepieux2021,Moskalets2014,Pekola2018,Sanchez2012,Karimi2021}. For simplicity, we consider first the tunneling as an example. Besides presenting the classical fluctuation-dissipation theorem for heat current, which we review in a general case after it, we also observe the intriguing quantum expression of heat current noise including the frequency dependent component due to zero point fluctuations surviving down to $T=0$. Next we focus on the temperature dynamics of a finite system coupled to a bath, which yields the experimentally accessible fundamental fluctuations of the effective temperature of this subsystem. Finally we review the experimental situation, with up to now only a small number of examples, on fluctuations in heat transport of quantum and classical systems. 

\subsection{FDT for heat in tunneling}\label{FDT tunneling}
We consider tunneling where the average heat current out from lead L was given by Eq.~\eqref{heattunn}. Taking for simplicity the normal conductors (NIN junction) with $n_L(\epsilon)=n_R(\epsilon)=1$, we have the average heat current at $eV=0$
\begin{equation}\label{QdotL2}
	\dot{Q}_L=\frac{1}{e^2R_T}\int d\epsilon~\epsilon[f_L(\epsilon)-f_R(\epsilon)].
\end{equation}
The thermal conductance for tunneling, $G_{\rm th}=\frac{d\dot{Q}_L}{dT_L}|_{T_L=T}$, is then given by $G_{\rm th}=\mathcal{L}TG_T$, where $G_T=1/R_T$ is the conductance of the tunnel junction.
Like the fully transmitting channels in Section~\ref{Thermoelectric transport}, the tunnel junction satisfies the Wiedemann-Franz law. 

The heat current operator $\dot {\hat H}_L$ to obtain the average heat current of Eq.~\eqref{heattunn} was calculated using the tunnel coupling operator of Eq.~\eqref{V} and commuting this with the Hamiltonian of the left lead. We may use this operator to find the two-time correlator of it and Fourier-transform to find the spectral density of noise of heat current at finite (angular) frequency $\omega$ (but at $eV=0$) as $S_{\dot Q}(\omega)=\int dt\langle \dot {\hat H}_L(t) \dot {\hat H}_L(0)\rangle e^{i\omega t}$ yielding~\cite{Averin2010,Sergi2011,Zhan2013,Karimi2021}
\begin{equation}\label{SQomega}
	S_{\dot Q}(\omega)=\frac{G_T}{6e^2}[(2\pi k_{\rm B}T)^2+(\hbar\omega)^2]\frac{\hbar\omega}{1-e^{-\hbar\omega/k_{\rm B}T}}.
\end{equation}
For the symmetrized noise, $S_{\dot Q}^{\rm (s)}(\omega)=\frac{1}{2}[S_{\dot Q}(\omega)+S_{\dot Q}(-\omega)]$, we then have
\begin{equation} \label{SQomegasymm}
	S_{\dot Q}^{\rm (s)}(\omega) =
	\frac{G_T}{12e^2}[(2\pi k_{\rm B}T)^2+(\hbar\omega)^2]\hbar\omega\coth(\frac{\hbar\omega}{2k_{\rm B}T}).  
\end{equation}
Now there are two important limits to consider. First of all, for $\omega \rightarrow 0$, we obtain the classical fluctuation-dissipation theorem for heat current as
\begin{equation} \label{FDTclass}
	S_{\dot Q}^{\rm (s)}(0) =2k_{\rm B}T^2 G_{\rm th}.  
\end{equation}
On the other hand, the finite frequency noise does not vanish at zero temperature, but
\begin{equation} \label{SQomegasymmT0}
	S_{\dot Q}^{\rm (s)}(\omega) =
	\frac{G_T}{12e^2}|\hbar\omega|^3, \,\,\, T=0.  
\end{equation}
\begin{figure}
	\centering
	\includegraphics [width=\columnwidth] {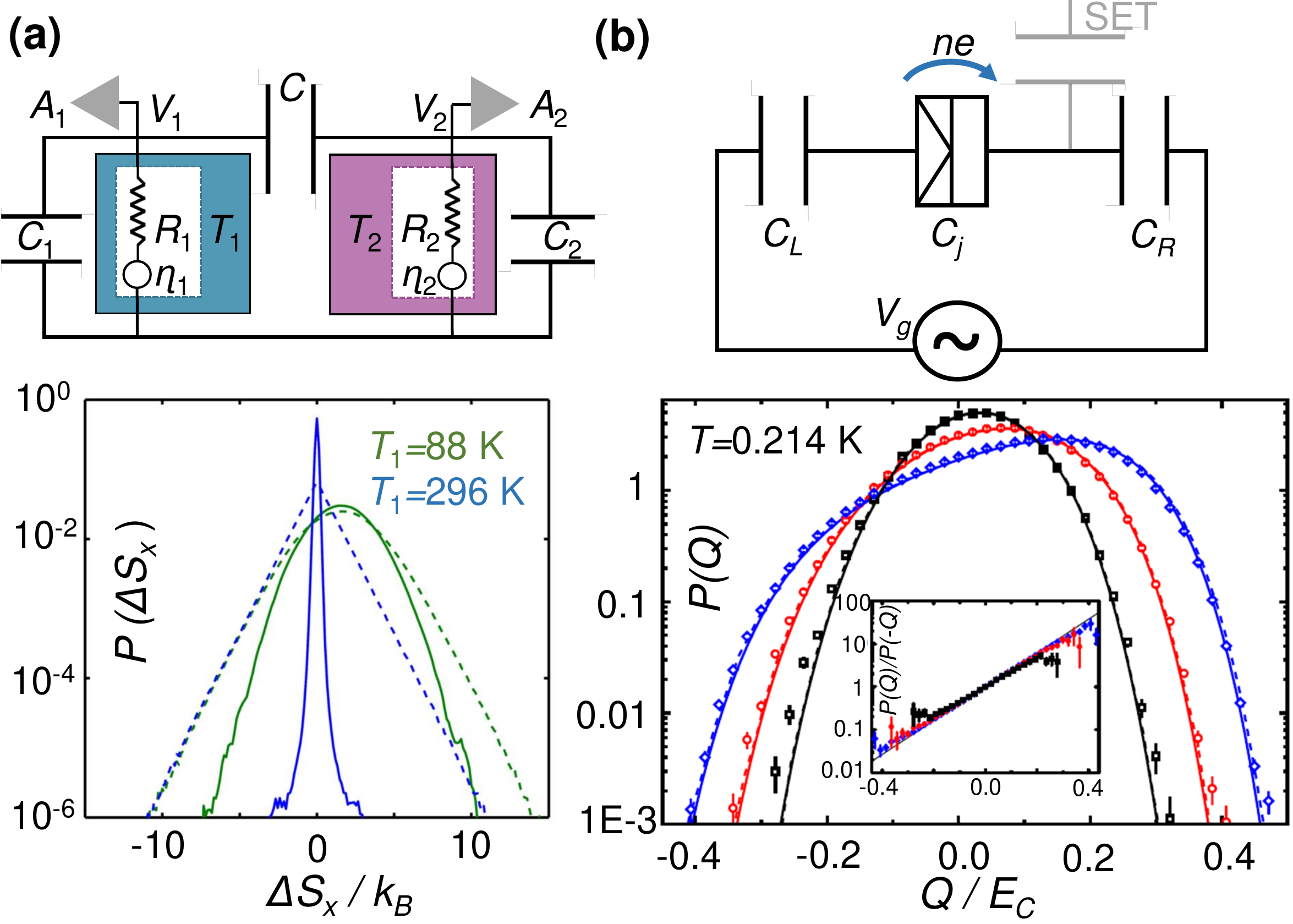}
	\caption{Indirect measurements of noise of heat current and entropy production. (a) Upper panel shows the circuit diagram where two resistors $R_1$ and $R_2$ at temperatures $T_1$ and $T_2=296$~K, respectively, are coupled via the capacitance $C$. $C_i$, $V_i$, and $A_i$ with $i=1,2$ show schematically the capacitance of the cables and input amplifiers, applied voltages, and low noise amplifiers, respectively. The probability of the entropy production due to the heat exchanged with the reservoirs $P(\Delta S_{r})$ (dashed lines) and the probability of the total entropy $P(\Delta S_{\rm tot})$ (solid lines) measured at different temperatures. Blue lines correspond to equilibrium where both distributions are centered symmetrically around zero and green lines are for out of equilibrium case with shifted distributions towards positive value. (b) In the upper panel, the electron box is formed by two metallic electrodes which are coupled by a tunnel junction with capacitance $C_j$, and connected capacitively to the voltage source via $C_L$ and $C_R$. The box is also connected to a single electron transistor (SET) working as an electrometer shown by gray line. The lower panel shows the measured distribution of the generated heat $P(Q)$ at three different drive frequencies: 1~Hz (black squares), 2~Hz (red circles), and 4~Hz (blue diamonds). The solid lines are exact theoretical predictions and dashed lines show the results of Monte Carlo simulations. Inset: The solid line displays the Crooks fluctuation theorem and the symbols show $P(Q)/P(-Q)$ ratio for the experimental distributions. (a) Adapted from~\cite{Ciliberto2013} and (b) adapted from~\cite{Saira2012a}.
		\label{CilibertoSaira}}
\end{figure}

\subsection{FDT for heat for a general system}\label{FDT general}
The previous sub-section serves as an illustration of how noise and dissipation are related. Here we extend the discussion to a general setup beyond the tunneling case. This would allow us to treat other mechanisms as well, for instance phonons, photons and electron-phonon coupling relevant for the current review. In general the FDT for heat applies in the form introduced in Eq.~\eqref{FDTclass} for low frequency noise. To see this we may write the Hamiltonian 
\begin{eqnarray}\label{E1}
	\mathcal{H}=\hat{H}_{\rm s}+\hat{H}_{\rm b}+\hat{H}_{\rm c}\equiv \hat{H}_0+\hat{H}_{\rm c},
\end{eqnarray}
where the unperturbed Hamiltonian $\hat{H}_0=\hat{H}_{\rm s}+\hat{H}_{\rm b}$ is composed of the system and bath and $\hat{H}_{\rm c}$ is again the coupling. Then in linear response, we have the expectation value of the heat current to the system
\begin{eqnarray}\label{E2}
	\dot{Q}=\langle \dot{\hat{H}}_{\rm s}\rangle=-\frac{i}{\hbar}\int_{-\infty}^{0}dt'\langle [\dot{\hat{H}}_{\rm s}(0),\hat{H}_{\rm c}(t')]\rangle_0,
\end{eqnarray}
The expectation value of a general operator $\mathcal{O}$ in the non-interacting system is written as $\langle \mathcal{O}\rangle_0=\Tr (e^{-\beta_{\rm s}\hat{H}_{\rm s}}e^{-\beta \hat{H}_{\rm b}}\mathcal{O})/\Tr (e^{-\beta_{\rm s}\hat{H}_{\rm s}}e^{-\beta \hat{H}_{\rm b}})$,
where $\beta_{\rm s}=1/k_{\rm B}T_{\rm s}$ and $\beta=1/k_{\rm B}T$ are the corresponding inverse temperatures of the system and bath, respectively. 
By definition, the thermal conductance is given by
\begin{eqnarray}\label{E4}
	G_{\rm th}=-\frac{d\dot{Q}}{dT_{\rm s}}|_{T_{\rm s}=T}=\frac{1}{k_{\rm B}T^2}\langle \delta \hat{H}_{\rm s}\dot{\hat{H}}_{\rm s}\rangle_0,
\end{eqnarray}
where we used $\hat{H}_{\rm s}-\langle \hat{H}_{\rm s}\rangle_0\equiv \delta \hat{H}_{\rm s}$. 
On the other hand, the spectral density of noise for the heat current at zero frequency is given by
\begin{eqnarray}\label{E12}
	S_{\dot{Q}}(0)=\int_{-\infty}^{\infty}dt'\langle \dot{H}_{\rm s}(t')\dot{H}_{\rm s}(0)\rangle_0
\end{eqnarray}
analogously to what was introduced in tunneling case. After some algebra and a careful comparison of Eqs.~\eqref{E4} and \eqref{E12} we find the FDT given in Eq.~\eqref{FDTclass}.

\subsection{Effective temperature fluctuations}\label{Effective temperature}
Here we consider a system with varying temperature $T(t)$. This setup, shown in Fig.~\ref{epset}(a), presents an absorber of a calorimeter or bolometer coupled via thermal conductance $G_{\rm th}$ to a heat bath at fixed temperature $T_0$. If we further assume that the small system has heat capacity $\mathcal{C}$, the energy balance equation reads for the heat current $\dot{Q}(t)$ between the bath and the absorber
\begin{equation}\label{dT}
	\dot{Q}(t)=\mathcal{C}\delta\dot{T}(t)+G_{\rm th}\delta{T}(t),
\end{equation}
where $\delta{T}(t)$ is the difference between the absorber temperature and that of the bath. In order to calculate thermal noise, we again evaluate the two-time correlator as
\begin{eqnarray}\label{correlator_2}
	\langle\dot{Q}(t)\dot{Q}(0)\rangle&&=\mathcal{C}^2\langle\delta\dot{T}(t)\delta\dot{T}(0)\rangle+G_{\rm th}^2\langle\delta{T}(t)\delta{T}(0)\rangle,
\end{eqnarray}
leading to
\begin{eqnarray}\label{noise7}
	S_{\dot{Q}}(\omega)=(\omega^2\mathcal{C}^2+G_{\rm th}^2)S_{T}(\omega).
\end{eqnarray}
Since we typically consider frequencies well below the temperature, $S_{\dot{Q}}(\omega)$ is essentially frequency independent, (which was shown by Eq. \eqref{SQomegasymm} for tunneling) and the classical FDT holds for $S_{\dot{Q}}(0)$ in form of Eq. \eqref{FDTclass} in equilibrium. Thus we have
\begin{eqnarray}\label{ST1}
	S_{T}(\omega)=\frac{2k_{\rm B}T_0^2}{G_{\rm th}}\frac{1}{1+(\omega\tau)^2},
\end{eqnarray}
where $\tau=\mathcal{C}/G_{\rm th}$ is the thermal relaxation time. This means that at very low frequencies $S_{T}(0)={2k_{\rm B}T^2}/{G_{\rm th}}$.
The root-mean-square (rms) fluctuation of temperature is obtained as inverse Fourier transform of the noise spectrum at $t=0$ as
\begin{eqnarray}\label{ST3}
	\langle \delta T^2\rangle&&=\int_{-\infty}^{\infty} \frac{d\omega}{2\pi}S_{T}(\omega)=\frac{k_{\rm B}T_0^2}{\mathcal{C}},
\end{eqnarray} 
which is the well-known textbook result of temperature fluctuations~\cite{Lifshitz,Berg_2015,Heikkilae2009}. The results of Eqs. \eqref{ST1} and \eqref{ST3} are directly accessible in experiments.

\subsection{Progress on measuring fluctuations of heat current and entropy}\label{noise Progress}
The previous discussion applies for systems and processes in or very near equilibrium. Over the past decades relations holding also far from equilibrium and for finite times have been developed~\cite{Bochkov1981,Jarzynski1997,Crooks1999,Seifert_2012}. During the past 20 years they have also become experimentally feasible mainly thanks to advances in production and manipulation of nanostructures. The best known non-equilibrium fluctuation relations governing entropy production $\Delta S$ are given by $P(\Delta S)/ P(-\Delta S)=e^{\Delta S/k_{\rm B}}$ and its corollary $\langle e^{-\Delta S/k_{\rm B}}
\rangle =1$. Here $\langle \cdot \rangle$ refers to the average over many experimental realizations or to the expectation value for the measurement. For macroscopic systems near equilibrium these relations simplify into second law of thermodynamics.

Here we give a brief summary of such non-equilibrium experiments on electrical systems. Fluctuations of entropy production and heat currents have been actively studied experimentally for more than a decade in the classical regime, but mainly via indirect means of detection since entropy is a tricky quantity for a direct measurement~\cite{Kleeorin2019}. Two main classes of systems under study have been those in the seminal experiments on molecules~\cite{Collin2005} and on electrical circuits~\cite{Kueng2012,Ciliberto2013,Saira2012a,Pekola2015,Berut2016}. These works go beyond FDT by addressing far-from-equilibrium fluctuation relations~\cite{Bochkov1981,Jarzynski1997,Crooks1999,Seifert_2012,Campisi2011,Pekola2019}. In the work by~\cite{Ciliberto2013} as shown in Fig.~\ref{CilibertoSaira}(a), the setup of two macroscopic resistors was examined at temperatures around the ambient. An indirect measurement of entropy was facilitated in~\cite{Ciliberto2013} via the detection of instantaneous electrical power $IV$, integrated over time and divided by the corresponding temperature of the macroscopic resistor. This way several fluctuation relations for entropy production under non-equilibrium conditions~\cite{Seifert_2012,Seifert2005} could be verified together with the standard FDT in linear response regime. Similarly in the setup of~\cite{Saira2012a} of Fig.~\ref{CilibertoSaira}(b), detecting single electrons making non-equilibrium transitions across a junction in a single-electron box provides indirect means of observing the dissipated energy and entropy production quantitatively~\cite{Averin2011,Koski2013}. These experiments were performed at temperatures three orders of magnitude lower than in~\cite{Ciliberto2013}. The Crooks~\cite{Crooks1999} and Jarzynski~\cite{Jarzynski1997} relations as well as generalized relations incorporating the role of information in Maxwell demon setup~\cite{Sagawa2010} could be tested accurately in these experiments~\cite{Pekola2019}.

The reason for using indirect measurement of heat by detailed electrical characterization is the fact that the powers are far too small to resolve, e.g. by direct thermometry (Section~\ref{Thermal Thermometry}). Next, we focus on progress of direct measurement of heat current fluctuations.

\begin{figure}
	\centering
	\includegraphics [width=\columnwidth] {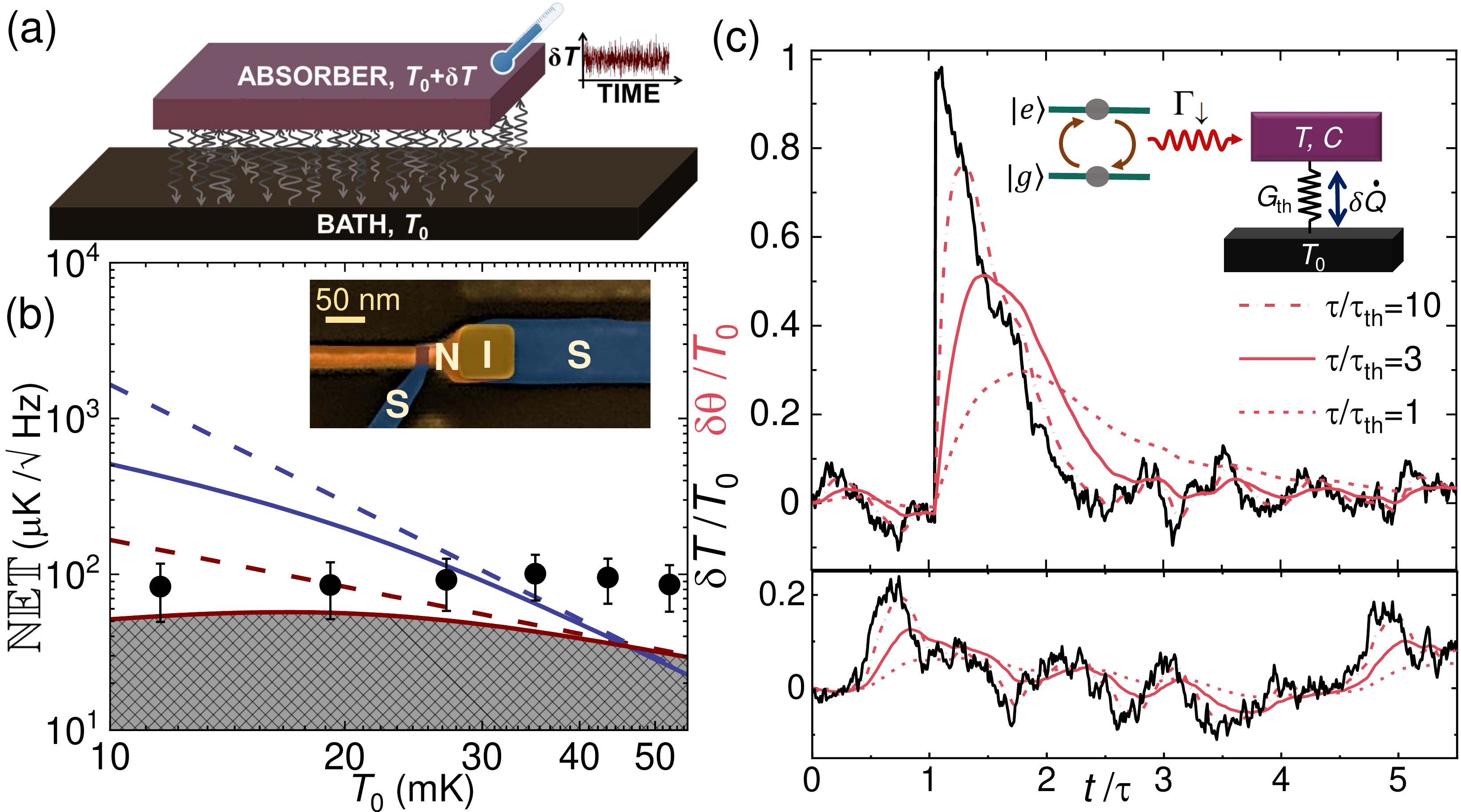}
	\caption{ Quantum calorimeter. (a) A normal-metal absorber is coupled to the phonon heat bath at fixed temperature $T_0$ via electron-phonon collisions, shown by many arrows. These collisions lead to stochastic exchange of heat between the absorber and the bath, and to fluctuating temperature $\delta T$ in the absorber. The core of the device is the calorimeter including a thermometer, which measures the temporal temperature variations. An example of a temperature trace is shown next to the absorber. (b) Noise equivalent temperature of the calorimeter. The solid symbols are the measured signal of temperature fluctuations in equilibrium $\mathbb{NET}\equiv \sqrt{S_T}=\sqrt{\langle \delta T^2\rangle/(2\Delta f)}$ obtained via the measurement of $\langle \delta T^2\rangle$ with $\Delta f=10$~kHz. The solid and dashed lines represent the noise-equivalent temperature in equilibrium $\mathbb{NET}$, of the normal-metal absorber in the presence and absence of an extra photon contribution, respectively. Red lines (the lower two) display $\mathbb{NET}$ in equilibrium $\mathbb{NET}_{\rm eq}=\sqrt{2k_{\rm B}T_0^2/G_{\rm th}}$ while the blue lines (the upper two) show $\mathbb{NET}=\delta \epsilon/\sqrt{\mathcal{C}G_{\rm th}}$, which is the required $\mathbb{NET}$ of a detector to observe a photon with energy $\delta\epsilon=1~{\rm K}\times k_{\rm B}$. The prohibited range bordered by the fundamental temperature fluctuations in equilibrium is indicated by the shaded area. (Inset) Scanning electron micrograph of part of the actual sensor, the SNIS structure, where Cu is used as a normal metal N and Al as a superconductor S. The SNIS junction is the dissipative element in the $RLC$ circuit operating at $f_0\approx 650$~MHz. (c) Simulation of the expected thermometer signal of a calorimeter in response to the absorption of an incoming photon depicted in the inset. The parameters of the simulation of the detector correspond to the experiment shown in (b), with $\eta\equiv\mathcal{C}/k_{\rm B}=100$ and $\hbar\omega_{\rm Q}/(k_{\rm B}T_0)=100$, where $\hbar\omega_{\rm Q}$ is the qubit energy, and $T_0=10$~mK. The noise seen in the traces is the result of temperature fluctuations due to the coupling of the absorber to the phonon bath. Upper panel shows a jump occuring at $t/\tau=1$ (the time instant set arbitrarily) exceeding the noise level of equilibrium fluctuations, with signal-to-noise ratio of about 10. The red lines (truncated by filtering) show the results at different cut-off frequencies of the thermometer, parametrized by the ratio of the electron-phonon time and the detector response time. (a) Adapted from~\cite{Karimi2020} and (b) adapted from~\cite{Karimi2020a}.
		\label{Temperature-fluctuations}}
\end{figure}

\subsection{Energy sensitivity of a calorimeter}\label{noise Energy}
The ultimate energy resolution of a thermal detector, see Fig.~\ref{Temperature-fluctuations}, is determined by the coupling of it to the heat bath associated with the fluctuations of heat current. Taking a wide band thermometer on a calorimeter, the rms fluctuations of the effective temperature due to this intrinsic noise are given by Eq.~\eqref{ST3}. In order to find the energy resolution of the detector one needs to compare this noise to the impact of the absorption of energy $E$ on the temperature of the detector, which can be evaluated by solving Eq.~\eqref{dT} for an instantaneous absorption of a photon at energy $E$ at time instant $t=0$, meaning $\dot Q(t)=E\delta(t)$ with the solution $\delta T(t)=(E/\mathcal C)e^{-t/\tau}\theta(t)$, where the time constant $\tau=\mathcal C/G_{\rm th}$ and $\theta(t)$ is the Heaviside step function. Thus at $t=0+$, the immediate rise of $T$ is $\delta T(0)=E/\mathcal C$. The signal-to-noise ratio $\mathbb{SNR}={\delta T(0)}/{\sqrt{\langle \delta T^2\rangle}}$ is
\begin{equation} \label{SNR}
	\mathbb{SNR} ={E}/{\sqrt{k_{\rm B}T_0^2 \mathcal C}},
\end{equation}
meaning that the energy resolution of the detector in this regime is 
\begin{equation} \label{energyresolution}
	\delta E=\sqrt{k_{\rm B}T_0^2 \mathcal C}.
\end{equation}
For convenience, let us write $\mathcal C = \eta k_{\rm B}$, where $\eta$ is a dimensionless constant that we assess below. Then we find that $\delta E =\sqrt{\eta} k_{\rm B} T_0$. As an example, related to the experiment of~\cite{Karimi2020}, we take a metallic calorimeter where $\mathcal C=\gamma \mathcal V T_0$ at low temperatures (see Fig.~\ref{Temperature-fluctuations}(a),(b)). Here $\gamma \sim 100 $ JK$^{-2}$m$^{-3}$ for copper and $\mathcal V < 10^{-21}$ m$^{3}$ is the volume of the absorber, yielding $\eta\sim 100$ and energy resolution $\delta E/k_{\rm B} \sim 0.1\,$K at $T_0=0.01\,$K~\cite{Karimi2020a}. 

The fluctuations in power have direct impact on the performance of calorimeters and bolometers~\cite{Irwin1995,Gildemeister2001}, i.e. thermal detectors of radiation. This noise determines the energy resolution of a calorimeter, and also the noise-equivalent power under continuous irradiation, like in the measurement of the cosmic microwave background~\cite{Mather1982}. Direct measurements of fluctuating temperature are rare (\cite{Chui1992,Karimi2020}). The pioneering measurement of \cite{Chui1992} employed a macroscopic calorimeter working at the so-called lambda point of liquid helium, i.e. its superfluid transition temperature at $T=2.17$~K. They managed to verify Eqs.~\eqref{ST1} and \eqref{ST3} thanks to a very high resolution of the thermometer measuring the magnetization of a paramagnetic salt (copper ammonium bromide) with the help of a SQUID down to $10^{-10}$~K$/\sqrt{\rm Hz}$ noise equivalent temperature.   

The nanofabricated detector of~\cite{Karimi2020} worked in the regime where the measurement cut-off frequency was 10~kHz, which falls somewhat below $1/\tau$. Furthermore the metallic absorber was proximitized by a superconducting contact further decreasing $\mathcal C$ and $G_{\rm th}$~\cite{Heikkilae2009a,Nikolic2020} and this way improving its performance. The experiment (Fig.~\ref{Temperature-fluctuations}(b)), utilizing a SNIS (superconductor-normal metal-insulator-superconductor) thermometer~\cite{Karimi2018} demonstrated noise of the effective temperature of the calorimeter that is very close to the expected fundamental fluctuation limit of Eq.~\eqref{ST1} at low frequencies, at the same time promising a SNR of $\sim 10$ in measuring an absorption event with photon energy $E/k_{\rm B}=1$~K. Figure~\ref{Temperature-fluctuations}(c) demonstrates by simulation the validity of the analysis above. The wide-band detector would present $T$ fluctuations which are an order of magnitude smaller than the temperature jump due to the 1~K photon absorption event. In summary of this measurement, it demonstrates the feasibility of a microwave photon measurement using a metallic calorimeter at $T_0=10\,$mK. 

There exist several other concepts of ultrasensitive thermal detectors, either metallic~\cite{Govenius2016,Kuzmin2019} or those utilizing graphene or semiconductors~\cite{Kokkoniemi2019,Kokkoniemi2020,Lee2020,LaraAvila2019,Roukes1999}, or those based on, e.g., temperature dependent magnetization~\cite{Enss2005,Kempf2018}. The advantage of graphene is its supposedly very low heat capacity that could make the thermal response time shorter than in metal detectors. Yet at the time of writing this paper, none of the proposed detectors has demonstrated detection of quanta in the said microwave regime. 

\section{Summary and outlook}\label{Summary}
In this review we have focused on fundamental aspects of quantum heat transport, moreover with main emphasis on experiments carried out during the past 20 years. It is probably fair to say that in many respects the physics of heat transport in quantum nanostructures is by now well understood, and experiments tend to confirm the theoretical predictions. In some systems clean experiments are, however, more difficult to realize than in others from practical point of view, and more experiments are needed: one example among others is presented by the one dimensional phonon structures where beautiful pioneering experiments were performed already long ago~\cite{Schwab2000}, but where precise conditions of how to realize ballistic contacts are still under debate. The experiments on quantum heat transport serve also as tools to understand quantum matter itself, like the recent experiments in the fractional quantum Hall regime demonstrate~\cite{Banerjee2017,Dutta2021}. On the other hand they show us ways of realizing new kinds of devices and of how to nail down and achieve their ultimate limits of performance. We discussed the latter issue in Section~\ref{Heat current noise} on noise in heat current. 

As to the potentially useful devices based on quantum heat transport, we discuss briefly two examples. The first one is a rather straightforward application of heat management on-chip for quantum information processes. Microwave photons provide a means to transport quanta and energy in general over large distances as we dicussed in Section~\ref{Photons Experiments}. It could thus serve as a way to reset quantum circuits rapidly. There is, however, a trade-off to be considered. Rapid thermalization is almost a synonym for low quality factor and fast decoherence of a quantum system, which are naturally not desirable properties. Therefore tunable coupling is a possible way to go, to switch on and off the coupling to a heat bath on demand. Variations of the many heat valves presented in this article could in principle serve the purpose. Tests of such an idea have been proposed and experimented on in~\cite{Partanen2018}. 

Quantum heat engines and cyclic refrigerators are presently under intensive study, see, e.g.~\cite{BENENTI20171,Humphrey2005,Alicki2018a,Campisi2016,Deffner2014,Brandner2017,Josefsson2018,Quan2007,Raja2021}. Experiments fully in the quantum regime are up to now practically nonexistent, although there are proposals that address realistic
setups~\cite{Karimi2016,Abah2012}. For instance, a so-called quantum Otto cycle can be realized by coupling a superconducting qubit alternately to two different heat baths~\cite{Karimi2016}. If this is done by varying the energy level separation of the qubit, like was done in the photonic heat valve or rectifier above, but now cyclically at RF frequencies, one can extract heat from the cold bath and dump it to the hot one, when system parameters are chosen properly. We expect devices of this type or analogous ones to work in the near future. Interesting questions arise on whether one can boost up the powers and/or efficiencies by exploiting quantum dynamics, and by what kind of protocols one can speed up the cycles for higher powers in general~\cite{Menczel2019,Funo2019,Solfanelli2020}.

We want to briefly note here that topological matter~\cite{Hasan2010,Qi2011}, specifically topological superconductors and Josephson junctions, have been proposed as potential novel elements in quantum thermodynamics and heat transport experiments due to their unconventional physical properties, see e.g., non exhaustive list of some recent work in~\cite{Rivas2017,Pan2021,Scharf2020,Bauer2019}. Due to the focus of the current paper, mainly on experiments, we do not discuss this topic further.  

In this review we have alluded to the connections of heat transport and quantum thermodynamics mainly regarding concrete device concepts, including thermal detectors, heat engines and refrigerators. On a more fundamental level, quantum heat transport is in the heart of open quantum systems physics~\cite{Breuer2002} with non-Hermitian dynamics governed by the quantum noise~\cite{Gardiner2010} widely discussed in this review. True thermodynamics counts on observations of heat currents and temperatures, and power consumption of the sources. Adopting this view, one can pose many questions like
how to measure work and heat in an open quantum system, for which the measurement apparatus cannot be viewed as an innocent witness of what is happening in the quantum system itself. The calorimeter can eventually become the microscope of quantum dynamics on the level of exchange of energy by individual quanta emitted or absorbed by the quantum system. This would give us the optimal tool to investigate stochastic thermodynamics in true quantum regime. Many other fundamentally and practically important questions arise and can potentially be answered by heat transport experiments. For instance, how does a quantum system thermalize, and does it find an equilibrium thermal state even in the absence of a heat bath? To conclude, investigations and exploitation of quantum heat transport will play an important role in the currently active field of quantum thermodynamics and in future quantum technologies in general.

\section{Acknowledgments}

This work was supported by Academy of Finland grant 312057, the European Union’s Horizon 2020 research and innovation programme under the Marie Sklodowska-Curie actions (grant agreement 766025), the Russian Science Foundation (Grant No. 20-62-46026), and Foundational Questions Institute Fund (FQXi) via Grant No. FQXi-IAF19-06.

\bibliography{RMP_JB_2021_temp}

\end{document}